\pdfoutput=1

\documentclass[3p,
               11pt]
               {elsarticle}
\usepackage{amssymb}

\usepackage{mySymbols}
\usepackage{myColors}
\usepackage{myPaperDefinitions}
\usepackage{multirow}
\usepackage{transparent}
\usepackage{cancel}
\newtheorem{remark}{Remark}

\usepackage{hyperref}
\hypersetup{%
	colorlinks,
	breaklinks=true,
	plainpages=false,%
	citecolor=blue,
	linkcolor=blue,
	urlcolor=blue,
	bookmarksopen=true,%
	bookmarksnumbered=false,%
	bookmarksdepth=5%
}

\begin{document}

\begin{frontmatter}



\title{
A consistent diffuse-interface finite element approach to rapid melt--vapor dynamics \replaced{with application to}{in} metal additive manufacturing
}

\author[SAM,TUM]{Magdalena Schreter-Fleischhacker\corref{cor1}}
\cortext[cor1]{Corresponding author}

 \author[SAM,TUM]{Nils Much}
 \author[TUB]{Peter Munch}
 \author[RUB]{Martin Kronbichler}
 \author[TUM]{Wolfgang A. Wall}
 \author[SAM,TUM]{Christoph Meier}

 \affiliation[SAM]{organization={Professorship of Simulation for Additive Manufacturing, Technical University of Munich},
 	addressline={Freisinger Landstraße 52}, 
 	city={Garching b. München},
 	postcode={85748}, 
 	country={Germany}
 }
 
 \affiliation[TUM]{organization={Institute for Computational Mechanics, Technical University of Munich},
             addressline={Boltzmannstrasse 15},
             city={Garching b. München},
             postcode={85748},
             country={Germany}}
             
\affiliation[RUB]{organization={Faculty of Mathematics, Ruhr University Bochum},
	addressline={Universitätsstraße 150}, 
	city={Bochum},
	postcode={44780}, 
	country={Germany}
}
\affiliation[TUB]{organization={Institute of Mathematics, Technical University of Berlin},
            addressline={Straße des 17. Juni 136}, 
            city={Berlin},
            postcode={10623}, 
            country={Germany}}

\begin{abstract}
	Metal additive manufacturing via laser-based powder bed fusion (\pbfam{}) faces perfor\-mance-critical challenges due to complex melt pool and vapor dynamics, often oversimplified by computational models that neglect crucial aspects, such as vapor jet formation. To address this limitation, we propose a consistent computational multi-physics mesoscale model to study melt pool dynamics, laser-induced evaporation, and vapor flow. In addition to the evaporation-induced pressure jump, we also resolve the evaporation-induced volume expansion and the resulting velocity jump at the liquid--vapor interface. We use an anisothermal incompressible Navier--Stokes solver extended by a conservative diffuse level-set framework and integrate it into a matrix-free adaptive finite element framework. To ensure accurate physical solutions despite extreme density, pressure and velocity gradients across the diffuse liquid--vapor interface, we employ consistent interface source term formulations developed in our previous work. These formulations consider projection operations to extend solution variables from the sharp liquid--vapor interface into the computational domain. Benchmark examples, including film boiling, confirm the accuracy and versatility of the model. As a key result, we demonstrate the model's ability to capture the strong coupling between melt and vapor flow dynamics in \pbfam{} based on simulations of stationary laser illumination on a metal plate. Additionally, we show the derivation of the well-known Anisimov model and extend it to a new hybrid model.	
This hybrid model, together with consistent interface source term formulations, especially for the level-set transport velocity, enables \pbfam{} simulations that combine accurate physical results with the robustness of an incompressible, diffuse-interface computational modeling framework.
\end{abstract}

%

\begin{keyword}


anisothermal evaporating two-phase flow \sep diffuse level-set framework \sep adaptive finite element method \sep melt--vapor interaction \sep laser-based powder bed fusion \sep metal additive manufacturing 
\end{keyword}

\end{frontmatter}



\section{Introduction}
\label{sec:intro}

\subsection{Background and challenges}
In additive manufacturing processes such as laser-based powder bed fusion additive manufacturing of metals (\pbfam{}), the use of high power lasers is very appealing due to the potential for exceptionally high build rates. \replaced{However, this approach also introduces complex physical phenomena, particularly in evaporation dynamics, which significantly influence part quality. These effects, summarized below, serve as the primary motivation for this contribution.}{However, this approach also introduces complex physical phenomena, particularly in evaporation dynamics, which can significantly influence part quality as summarized in the following and serves as primary motivation of this contribution.} 


The physical impact of evaporation during the manufacturing process on melt pool thermo-hydrodynamics has been discussed in the review articles~\cite{debroy2018additive,meier2017thermophysical} and experimentally studied in various works~\cite{matthews2016denudation,zhou2015balling,zhirnov2018evaporation,bidare2018fluid,cunningham2019keyhole,zhao2019bulk,leung2019effect}, among others. Specifically in \pbfam{}, the energy absorbed by the powder bed leads to extreme peak temperatures exceeding the boiling point, causing rapid evaporation. The high density change during the evaporative phase change induces a recoil pressure that acts as the dominant excitation force on the melt pool surface, resulting in deep keyhole depressions within the melt pool~\cite{cunningham2019keyhole}. This effect is further amplified by increased laser energy absorption through multiple reflection-absorption events at the keyhole walls. The unstable coupling between recoil pressure, thermocapillary forces, keyhole morphology and laser absorption can lead to periodically oscillating and collapsing melt pool depressions, leaving performance-critical subsurface pores in the rear end of the melt pool~\cite{brennan2021defects}. The interplay between recoil pressure and vapor/gas flow further contributes to negative effects such as
undesired spattering of melt drops, powder particle ejection~\cite{zhao2019bulk}, as well as denudation (clearing of powder around a single-track bead), resulting in \deleted{empty} powder-free denudation zones due to lateral entrainment of powder particles into the melt pool~\cite{matthews2016denudation,bidare2018fluid,qiu2015role,ly2017metal, bitharas2022interplay}. 
In addition, the resulting vapor jet absorbs and scatters the incident laser radiation and acts as a means of transport for potential pollutants~\cite{bauerle2013laser}. While the increased energy absorption and
melting depth in keyhole mode enable very high build rates, the aforementioned effects collectively contribute to excessive porosity and surface roughness, forming potential crack initiation sites for fatigue failure and thus significantly affecting the final part quality. 

To enhance understanding of the governing physical processes of melt pool thermo-hydrodynamics on the mesoscale, i.e., on the order of micrometers, physics-based computational models have been proposed. These models, as outlined in review articles~\cite{debroy2018additive,meier2017thermophysical,megahed2016metal,markl2016multiscale,cook2020simulation}, play a central role in establishing a simulation-based characterization of metal additive manufacturing processes.
A key component of these models is the accurate modeling of melt--vapor interaction, which is the specific focus of the present work.


\subsection{Related work on unresolved and resolved evaporation models in computational melt pool simulations}

Among existing thermo-hydrodynamic melt pool models, we distinguish between \emph{unresolved} and \emph{resolved} evaporation models in the following: 

In \emph{unresolved} evaporation models, the flow in the ambient gas phase is neglected, although experimental findings~\cite{ly2017metal, bitharas2022interplay} demonstrate that vapor-driven entrainment of particles by ambient gas flow has been determined as the governing mechanism for powder particle movement. Evaporation in these models is typically considered through simplified models that solely account for the evaporation-induced pressure jump~\cite{anisimov1995instabilities} and evaporative cooling. 
Such models were established in the finite-element(FEM)-based multiphysics framework ALE3D~\cite{noble2017ale3D} from the Lawrence Livermore Laboratory, e.g.~\cite{matthews2016denudation,khairallah2016laser,khairallah2014mesoscopic,ly2017metal,martin2019dynamics}.
Other mesoscale models include~\cite{lee2015mesoscopic}, using the \replaced{finite volume}{finite difference} method based on the code FLOW-3D,
and~\cite{korner2013fundamental,korner2011mesoscopic,ammer2014simulating} using the Lattice-Boltzmann method. 
Simplified phenomenological evaporation models were employed in~\cite{leitz2018fundamental} using the available phase-field FEM framework in COMSOL and in~\cite{qiu2015role,panwisawas2017mesoscale,geiger20093d} using OpenFOAM. 
Smoothed particle hydrodynamics (SPH) models were presented in~\cite{weirather2019smoothed,meier2020meltpool,fuchs2022versatile}. SPH is an effective numerical technique to represent multiphase problems with phase transitions, but is not deemed to be suitable for evaporative phase changes involving significant density changes.

The \emph{resolved} modeling of the vapor phase considers the exchange of mass, the resulting pressure jump and flow of vapor, and the convective heat transfer caused by the vapor/gas velocity. These aspects, along with their combined effects on the dynamics of the melt pool, are crucial for accurately describing the physics of \pbfam{} on the mesoscale. 
However, this approach introduces computational challenges due to the presence of strong discontinuities in velocity, pressure and density at the liquid--vapor interface. To the best of our knowledge only a few attempts have been made in the literature to model these effects comprehensively: In the context of welding,~\cite{tan2013investigation, tan2014analysis, kouraytem2019effect} developed a model with explicitly resolving the compressible vapor/gas phases and applying the evaporation-induced pressure jump at the interface phenomenologically as an interface jump condition.
A framework for modeling directed energy deposition (DED) processes, combining an incompressible Navier--Stokes solver with a diffuse level-set method, was presented in~\cite{lin2020conservative, zhu2021mixed}, utilizing the local fluid velocity for level-set transport followed by a mass-fixing procedure. \added{
Flint et al.~\cite{flint2022fundamental, flint2020thermal, flint2023fundamental} developed a multi-phase thermal incompressible fluid dynamics model using the Volume of Fluid method in OpenFOAM to study evaporation/condensation in multi-component substrates during laser-induced melting. They accounted for evaporation-induced dilation at the interface and incorporated a source term for phase change effects in the transport equations.}
The commercial software COMSOL was used in~\cite{courtois2014complete} to simulate keyhole formation in laser welding, explicitly resolving the vapor phase, and, in~\cite{mayi2019laser} to study the dependence between background gas properties and particle entrainment through the vapor plume based on a simplified ALE-based axisymmetric model.
In a recent contribution by Zenz et al.~\cite{zenz2024compressible}, a mass and energy conservative compressible multiphase flow model including phase changes was developed to study keyhole dynamics in metal additive manufacturing and welding. They used a variant of the volume-of-fluid method, implemented it using the finite volume method into OpenFOAM and obtained good agreement between the numerical results and the experiments by Cunningham et al.~\cite{cunningham2019keyhole}. 

\replaced{A high-fidelity predictive model of melt pool thermo-hydrodynamics---accurately capturing molten metal fluid dynamics, laser-induced evaporation, and the resulting vapor and ambient gas flow---remains an ongoing challenge in \pbfam{} and related processes.
While significant progress has been made using an incompressible flow framework, as demonstrated by Flint et al.~\cite{flint2024version}, and a compressible flow framework, as shown by Zenz et al.~\cite{zenz2024compressible}, no single model yet integrates all relevant physical phenomena, such as modeling the effects of mass-conserving evaporative phase change, laser-material interaction, multi-species diffusion, and gas compressibility in a unified framework.
In this work, we bridge a key gap in the development of robust incompressible flow models with resolved vapor phase modeling by incorporating new formulations for evaporative phase change, crucial for accurate prediction of the evaporated mass and melt-vapor dynamics. Additionally, we propose a hybrid recoil pressure model that aligns the resulting recoil pressure with gas dynamics relations in~\cite{anisimov1995instabilities}. To the best of our knowledge, these advancements have not been achieved in previous melt pool models for \pbfam{} and represent a crucial step toward a more predictive and physically consistent simulation framework.}{Summarizing, a high-fidelity predictive model of the melt pool thermo-hydrodynamics, accurately capturing the fluid dynamics of the liquid metal, laser-induced evaporation as well as the resulting vapor and ambient gas flow is of major importance, but is still pending in the field of melt pool modeling in \pbfam{}. }

\subsection{Our contributions}

We present a new consistent computational multi-physics model for studying the melt--vapor interaction in \pbfam{} processes. Central to our model is the consideration of the coupled anisothermal incompressible fluid dynamics of molten metal, laser-induced evaporation, and subsequent vapor flow. We use an adaptive finite element framework and a diffuse representation of the liquid--vapor/gas interface, chosen for its inherent robustness in handling the complex rapid and drastic changes in interface topology due to evaporation dynamics. To ensure accurate, physically meaningful solutions despite extreme density, pressure and velocity gradients across the finite liquid--vapor interface, we incorporate new mathematically consistent formulations for computing regularized interface source terms and the level-set transport velocity, developed in our previous research works~\cite{schreter2024consistent,much2023}. We outline the main objectives of this study as twofold:

\begin{sloppypar}
	First, building on our previously proposed isothermal level-set-based diffuse-interface framework for evaporating two-phase incompressible flows~\cite{schreter2024consistent}, we extend the model to anisothermal conditions by \emph{integrating a two-phase heat transfer model and introducing temperature-dependent interface fluxes and forces}. 
	\replaced{Additionally, in a theoretical study, we propose a refined formulation for the level-set transport velocity by incorporating a curvature correction term into our velocity projection method for the liquid or gas phase onto the finite interface region~\cite{schreter2024consistent}. This refinement improves the accuracy of predicting evaporated mass and interface topology and is worth investigating in future computations once an efficient extension algorithm is available.}{
	In addition, we propose a refined formulation for the level-set transport velocity. Thereto, building on our velocity projection method for the liquid or gas phase onto the finite interface region introduced in~\cite{schreter2024consistent}, we enhance the approach by incorporating a curvature correction term. 
	This approach has the potential to further improve the accuracy of predicting evaporated mass and interface topology.}
\end{sloppypar}	
	Second, we enhance the anisothermal evaporating two-phase flow framework to a mathematically consistent computational multi-physics model to study melt pool dynamics. To this end, 
	we introduce a \emph{new resolved formulation for an evaporation model} suitable for an incompressible multi-phase flow framework, using well-established relations for the evaporative mass flux and the evaporation-induced recoil pressure~\cite{knight1979theoretical,anisimov1995instabilities}. 
	In this context, a derivation of the well-known recoil pressure model according to Anisimov et al.~\cite{anisimov1995instabilities} is shown and extended to yield a novel hybrid model. This hybrid model together with the proposed regularized interface flux and level-set transport velocity formulations allow for \pbfam{} simulations combining accurate physical results with the robustness of an incompressible, diffuse-interface computational modeling framework.
	In addition, we consider an immobile, rigid solid phase and account for solid--liquid phase transitions of melting and solidification. 
	
 Besides, we leverage exascale high-performance computing techniques, such as matrix-free operator evaluation~\cite{kronbichler2012generic,Kronbichler18multiphase,proell2023highly,munch2023,proell2024highly}, adaptive mesh refinement together with available parallelized MPI-based
 implementations using domain decomposition from the open-source finite element library \texttt{deal.II}~\cite{africa2024deal}. These techniques are considered crucial to enhance computational efficiency for the present complex multi-physics problem. In addition, we use and have extended the  open-source incompressible Navier--Stokes solver \texttt{adaflo}~\cite{Kronbichler18multiphase}. Details on the numerical framework can be found in our previous work~\cite{schreter2024consistent}. \added{The source code and input files required to reproduce the
 numerical experiments are available online in the reproducibility repository~\cite{schreter2025zenodo}.}

The remainder of this article is organized as follows: The governing equations of the anisothermal two-phase flow framework with evaporative phase change, including the refined formulation of the level-set transport velocity 
are presented in Section~\ref{sec:two_phase_flow}. In Section~\ref{sec:numerical_examples_two_phase_flow}, benchmark examples of pure anisothermal two-phase flows are presented to verify our computational framework by comparison with analytical solutions and/or reference results from the literature. In Section~\ref{sec:melt_pool}, we present the mathematical framework of the mesoscale melt pool model, including our proposal of the resolved evaporation model. In Section~\ref{sec:numerical_examples_pbfam}, we demonstrate the application of the melt pool model to study melt--vapor dynamics during stationary laser illumination in \pbfam{} and evaluate the strengths and weaknesses of the model using 1D to 3D simulations. Conclusions are drawn in Section~\ref{sec:conclusion}.

\section{An anisothermal incompressible two-phase flow framework with evaporative phase change}
\label{sec:two_phase_flow}

In the following, we present a diffuse model for anisothermal two-phase flow with evaporative phase change.  This model extends our previous work on isothermal two-phase flow with evaporative phase change (detailed description \replaced{provided}{can be found} in~\cite{schreter2024consistent}) to account for anisothermal conditions.  
Our mathematical model builds upon the following assumptions for the present study:
\begin{itemize}
	\item Incompressible, viscous (Newtonian) flow at moderate Reynolds numbers.
	\item Convective and conductive heat transfer, neglecting contributions due to viscous dissipation. 
	\item Diffuse interface transition region with a finite but small interface thickness for thermophysical properties and interface fluxes between the phases.
	\item Spatially resolved model for the vapor phase and the liquid--vapor phase transition to explicitly
	model the dynamics through evaporation-induced recoil pressure and vapor flow.
\end{itemize}
Our Eulerian domain of interest $\Omega=\OmegaG\cup\OmegaL\in\mathbb{R}^{n}$ with $n\in\{1,2,3\}$ is divided into a liquid phase $\OmegaL$ and a vapor/gas phase $\OmegaG$. Irreversible evaporation of the liquid phase \deleted{may} occur\added{s} along the liquid-gaseous interface $\Gamma\in\mathbb{R}^{n-1}$ \added{in the presence of an evaporative mass flux}.
By employing a level-set based diffuse interface capturing scheme for determining the position of the liquid--vapor interface, a single set of equations for the two-phase flow domain \replaced{is obtained}{can be derived}.

\subsection{Anisothermal incompressible Navier--Stokes equations}

The velocity field $\boldsymbol{u}(\Bx,t)$, the pressure field $p(\Bx,t)$ and 
the temperature field $T(\Bx,t)$ for point $\Bx\in\Omega$ and at time~$t\,\in[0,\tEnd]$ are governed by the incompressible, anisothermal Navier--Stokes equations formulated in an Eulerian setting. They consist of the continuity equation, the momentum balance equation and the energy equation:
\begin{subequations}
	\begin{align}
		\label{eq:continuity}	
		\nabla\cdot\boldsymbol{u} &= \evaporDilationRate
		\quad&&\text{ in }\Omega\times[0,\tEnd]\,,\\
		\label{eq:momentum_balance}
		\rhoEff\left(\fracPartial{\boldsymbol{u}}{t}
		+(\boldsymbol{u}\cdot\nabla)\,\boldsymbol{u}
		\right) &= -\nabla p + \diver\viscousStress+\rhoEff\,\boldsymbol{g}+
		\boldsymbol{f}
		&&\text{ in }\Omega\times[0,\tEnd]\,,\\
		\label{eq:heat_transfer}
		\rhoCpEff\left(\fracPartial{T}{t} + \boldsymbol{u}\cdot\nabla T\right) &= \nabla\cdot{\left(\kEff\,\nabla T\right)} + s
		\quad&&\text{ in }\Omega\times[0,\tEnd]\,.
	\end{align}
\end{subequations}
Equations  \eqref{eq:continuity}-\eqref{eq:heat_transfer} are supplemented by suitable initial conditions
\begin{equation}
	\Bu=\Bu^{(0)}, T=T^{(0)} \quad\text{in }\Omega\times\{t=0\}
\end{equation}
where the superscript $(\bullet)^{(0)}$ indicates an initial field function.
Dirichlet and Neumann boundary conditions are imposed according to
\begin{align}
	\Bu &= \bar{\Bu}&&\text{on }\partial\Omega_{\text{D},u}\subset\partial\Omega\times[0,\tEnd], \\
	\boldsymbol{\sigma}\cdot\normalDomain&= \bar{\Bt}\text{ if }\Bu\cdot\hat{\Bn} > 0 \text{ (outflow)}&&\text{on }\partial\Omega_{\text{N},u}\subset\partial\Omega\times[0,\tEnd],\\
	T &= \bar{T}&&\text{on }\partial\Omega_{\text{D},T}\subset\partial\Omega\times[0,\tEnd],\\ \Bq\cdot\hat{\Bn} &= \bar{q}&&\text{on }\partial\Omega_{\text{N},T}\subset\partial\Omega\times[0,\tEnd]
\end{align}
with the Cauchy stress tensor $\boldsymbol{\sigma}=\viscousStress - p\,\mathcal{I}$, where $\mathcal{I}$ denotes the second-order identity tensor. For the viscous stress tensor, we use the formulation proposed in~\cite{schreter2024consistent},  $\viscousStress=2\,\muEff\left(\mathbb{\varepsilon}-\Tr\left(\mathbb{\varepsilon}\right)\,\nGamma\otimes\nGamma\right)$. Here, we subtract the non-physical volumetric deformation $\Tr\left(\mathbb{\varepsilon}\right)$ along the interface normal direction $\nGamma$, which is caused by the introduction of the diffuse evaporative dilation rate $\evaporDilationRate$ in the continuity equation ($\Tr\left(\mathbb{\varepsilon}\right)\equiv\diver\Bu=\evaporDilationRate$, cf. \myeqref{eq:continuity}). Due to the regularized formulation for the evaporative dilation rate $\evaporDilationRate$, explained in Section~\ref{sec:interface_fluxes}, the corrective term  is only non-zero in the finite interface region. 
The heat flux is considered as $\Bq=-\kEff\,\nabla T$ according to Fourier's law. The outward-pointing unit normal vector to the domain boundary $\partial\Omega=\partial\Omega_{\text{D},u}\cup\partial\Omega_{\text{N},u}=\partial\Omega_{\text{D},T}\cup\partial\Omega_{\text{N},T}$ is denoted as $\normalDomain$  with $\partial\Omega_{\text{D},u}\cap\partial\Omega_{\text{N},u}=\partial\Omega_{\text{D},T}\cap\partial\Omega_{\text{N},T}=\emptyset$.

The thermophysical properties, indicated by the subscript $\eff{(\bullet)}$, are assumed to be constant\added{, i.e. temperature-independent,} within the (pure) liquid or vapor phases and exhibit a smooth transition across the finite but small interface region. They comprise the effective density $\rhoEff$, the effective dynamic viscosity $\muEff$, the effective volume-specific heat capacity $\rhoCpEff$ and the effective thermal conductivity $\kEff$, which are specified in Section~\ref{sec:matProps}.

Gravitational forces are considered by $\eff{\rho}\,\boldsymbol{g}$. Additional forces in the momentum balance equation are denoted by $\boldsymbol{f}$, while additional fluxes in the energy equation are denoted by $s$. These additional contributions are specified in Section~\ref{sec:interface_fluxes} and in Section~\ref{sec:source_terms_melt_pool}. 

\subsection{Level-set framework}
\label{sec:level_set}

The temporal evolution of the liquid--vapor interface $\GammaLevelSet$ is represented by the zero-isosurface of a regularized level-set function $-1\leq\phi(\boldsymbol{x},t)\leq1$ using the level-set framework by Olsson et al.~\cite{olsson2007conservative}. We denote $\phi>0$ as inside the liquid phase and $\phi<0$ as inside the gaseous phase.  The temporal evolution of the level-set function $\phi$ is obtained by solving the advection equation
\begin{align}
	\label{eq:transport}
	\fracPartial{\phi}{t}+\uGamma\nabla\phi&=0  \quad\text{ in }\Omega\times[0,\tEnd]\,
\end{align}
with the level-set transport velocity $\uGamma$, for which mathematically consistent formulations to accurately predict the evaporated mass and the resulting interface topology are discussed in Section~\ref{sec:level_set_transport_velocity}.
Equation \eqref{eq:transport} is supplemented by the initial condition 
\begin{equation}
	\label{eq:initial_phi}
	\phi^{(0)} = 
	\tanh\left(\frac{d(\Bx)}{2\epsilon}\right)\quad\text{in }\Omega\times\{t=0\}\,
\end{equation}
depending on the interface thickness parameter $\epsilon$ and \replaced{the}{a} signed distance function $d(\Bx)$. 
\begin{remark}
\added{The interface thickness parameter $\epsilon$ governs the resolution of the level-set field and the accuracy of phase transition and interface fluxes by influencing the smoothed Heaviside and consequently delta functions (see Sections~\ref{sec:matProps} and \ref{sec:interface_fluxes}). Following~\cite{olsson2007conservative}, we set $\epsilon$ proportional to the mesh size in the interface region to ensure a well-resolved transition zone. According to~\cite{olsson2007conservative,much2023}, resolving the interface with 8–32 elements minimizes the spatial discretization error. Too small values cause spurious oscillations in the solution, while too large values introduce artificial diffusion and reduce accuracy. However, achieving accurate melt pool dynamics in \pbfam{} simulations demands particularly small interface thicknesses and correspondingly fine mesh resolutions, as discussed in~\cite{much2023} and summarized in the following. Achieving a 1\% error in the recoil pressure compared to a sharp-interface model requires an interface thickness of $w_\Gamma\approx\SI{0.1}{\micro\meter}$
	when using a diffuse two-phase heat transfer model with representative \pbfam{} parameters. The interface thickness is calculated from the interface thickness parameter $\epsilon$ as $w_\Gamma=6\,\epsilon$.
While coarser interface thicknesses introduce only negligible temperature deviations compared to a sharp-interface model across the domain, they cause substantial recoil pressure errors about one order of magnitude higher due to its exponential dependence on temperature. Consequently, for accurate diffuse-interface melt pool models more
efficient implementations and the use of high-performance computing infrastructures at a larger scale are required.}
\end{remark}
	
Dirichlet and Neumann boundary conditions are prescribed
\begin{align}
	\phi&=\bar{\phi}\text{ on  }\partial\Omega^{\text{inflow}}_{\text{D},\phi}\subset\partial\Omega\times[0,\tEnd]\,,\qquad \nabla\phi\cdot\hat{\Bn}=0\text{ on  }\partial\Omega_{\text{N},\phi}\subset\partial\Omega\times[0,\tEnd]
\end{align}
along the domain boundary $\partial\Omega=\partial\Omega^{\text{inflow}}_{\text{D},\phi}\cup\partial\Omega_{\text{N},\phi}$ with $\partial\Omega=\partial\Omega^{\text{inflow}}_{\text{D},\phi}\cap\partial\Omega_{\text{N},\phi}=\emptyset$.

To keep the shape of the level-set function profile constant as the interface moves, reinitialization is additionally performed according to~\cite{olsson2007conservative}. From the level-set field, the interface normal vector $\nGamma$ and the mean curvature $\kappa$ are derived by performing a filtering procedure~\cite{olsson2007conservative,schreter2024consistent}. \added{For } details on the numerical solution procedure for the level-set framework \replaced{refer to}{can be found in}~\cite{schreter2024consistent}.

\subsection{Effective thermophysical liquid-gas properties}
\label{sec:matProps}
From the level-set function $\phi$, a localized, indicator-like representation \replaced{is}{can be} constructed by employing the smoothed approximation of the Heaviside function~\cite{sussman1994level,peskin2002immersed}
\begin{equation}
	\label{eq:heaviside}
	H_\phi(\phi) = 
	\begin{cases} 	
		0 & d(\phi) \leq -3\epsilon\\
		\frac{1}{2} + \frac{d(\phi)}{6\,\epsilon} + \frac{1}{2\,\pi}\,\sin\left(\frac{\pi\,d(\phi)}{3\,\epsilon}\right) & -3\epsilon < d(\phi) < 3\epsilon\\
		1 & d(\phi) \geq 3\epsilon
	\end{cases} \,.
\end{equation}
This function is used to interpolate quantities between the two phases. For example, the effective dynamic viscosity $\muEff$, the effective thermal conductivity $\kEff$ and the effective volume-specific heat capacity $\rhoCpEff$ -- grouped together as $\eff{a}$ --
 are evaluated as arithmetic phase-fraction weighted average of the values for the liquid ($\liqP{a}$) and the gaseous phase ($\gasP{a}$)
\begin{equation}
	\eff{a}\left(\phi\right) = H_\phi(\phi)\,\liqP{a}
	+ \left(1-H_\phi\left(\phi\right)\right)\,\gasP{a}\, \text{ for } a \in \{\mu, k, \rho\cp \}.
	\label{eq:muEff}
\end{equation}
For obtaining consistency with the expression  of the evaporative dilation rate (see Section~\ref{sec:interface_fluxes}) and satisfying local mass conservation (see discussion in~\cite{schreter2024consistent}), we employ a reciprocal interpolation function of the density between the two phases
\begin{equation}
	\frac{1}{\func{\rhoEff}{\phi}} = \frac{\Hphi(\phi)}{\rhoL} + \frac{1-\Hphi(\phi)}{\rhoG}
	\label{eq:rhoEff}
\end{equation}
considering the density of the liquid phase $\rhoL$ and the one of the vapor phase $\rhoG$. \added{We note that although we employ a reciprocal interpolation function for the effective density, we have chosen to use arithmetic phase-fraction-weighted averages for the volume-specific heat capacity $\rhoCpEff$ to achieve the highest accuracy in our diffuse interface model compared to the sharp interface model, supported by the findings in~\cite{much2023}.}

\subsection{Interface source terms}
\label{sec:interface_fluxes}
For the present anisothermal incompressible vaporizing two-phase-flow model~with the governing equations presented in \myeqrefs{eq:continuity}{eq:heat_transfer}, the two phases are coupled by interface source terms consisting of (i) the evaporative dilation rate $\evaporDilationRate$, (ii) the temperature-dependent surface tension force $\boldsymbol{f}=\surfaceTensionForce$, and (iii) the evaporative cooling $s=\evaporCooling$. We employ a regularized representation of the singular interface contributions over a finite but small thickness of the interface region, denoted by a superscript tilde $\tilde{(\bullet)}$, by considering a parameter-scaled smoothed Dirac delta function, as we proposed in~\cite{much2023}. The latter is constructed from the symmetric, smoothed Dirac delta function
\begin{equation}
	\label{eq:delta}
	\symDelta(\phi)=||\nabla H_\phi(\phi)||\,
\end{equation}
using the smoothed Heaviside function \eqref{eq:heaviside}. In the following we abbreviate the spatial and temporal dependencies of the primary variables as $\phi:=\phi(\Bx,t)$ and $T:=T(\Bx,t)$.

\paragraph{Evaporative dilation rate}

We consider a regularized formulation of the evaporation-induced volume expansion and consequently the
evaporation-induced velocity as well as pressure jumps across the finite liquid--vapor interface region. The regularized representation of the evaporative dilation rate reads
\begin{equation}
	\label{eq:evaporDil}
	\func{\evaporDilationRate}{\Bx,t}=\func{\mDot}{T}\,\left(\frac{1}{\rhoL}-\frac{1}{\rhoG}\right)\func{\symDelta}{\phi}
\end{equation}
and enters the continuity equation~\eqref{eq:continuity}. Compared to our previous work~\cite{schreter2024consistent}, where we prescribed an analytical, temperature-independent function for the evaporative mass flux  $\func{\mDot}{T}$, for anisothermal conditions as considered in the present work, the evaporative mass flux needs to be computed from the temperature field. There are several models available for calculating the evaporative mass flux. At this point, we do not restrict ourselves to a specific model, but leave its choice to respective applications presented in Sections~\ref{sec:numerical_examples_two_phase_flow} and \ref{sec:numerical_examples_pbfam}. 

\paragraph{Temperature-dependent surface tension}

We employ a regularized interface force formulation~\cite{brackbill1992} for the temperature-dependent surface tension force including thermal Marangoni effects, leading to the force contribution in the momentum equation~\eqref{eq:momentum_balance}
\begin{equation}
	\label{eq:surface_tension}
	\func{\surfaceTensionForce}{\Bx,t}=\left(\surfTenCoeff(T)\,\kappa(\phi)\nGamma(\phi)
	+ \left(\mathcal{I} - \nGamma(\phi) \otimes \nGamma(\phi)\right) \nabla \alpha(T) \right) \func{\deltaRho}{\phi}\,.
\end{equation}
The potentially temperature-dependent surface tension coefficient reads
\begin{equation}
	\surfTenCoeff(T) = \max\left(\surfTenCoeffConst - \surfTenCoeffTemp\left(T-\TRefSurfTen\right), \surfTenCoeffRes\right)
\end{equation}
considering the surface tension coefficient $\surfTenCoeffConst$ at reference temperature $\TRefSurfTen$, the surface tension gradient coefficient $\surfTenCoeffTemp$ and a residual surface tension coefficient for numerical reasons set to $\surfTenCoeffRes=0.01$. As a special case, temperature-independent surface tension \replaced{is}{can be} modeled by setting $\surfTenCoeffTemp$ to zero, resulting in $\surfTenCoeff=\surfTenCoeffConst$ and $\nabla\surfTenCoeff=\boldsymbol{0}$. The interface mean curvature $\kappa$ and the interface normal vector $\nGamma$ 
are computed from a filtering procedure of the level-set function, as described in~\cite{schreter2024consistent}. The parameter-scaled delta function $\deltaRho$ of our previous work~\cite{much2023}
\begin{equation}
	\func{\deltaRho}{\phi} = \func{\symDelta}{\phi}\,\func{\rhoEff}{\phi}\,c_\rho\quad\text{ with }c_\rho = \frac{\gasP{\rho}-\liqP{\rho}}{\liqP{\rho}\gasP{\rho}\ln\left(\frac{\gasP{\rho}}{\liqP{\rho}}\right)} \text{ for } \gasP{\rho} > 0 \wedge \liqP{\rho} > 0\,. 
\end{equation}
is employed to achieve increased solution accuracy compared to the classical CSF model using the non-parameter-scaled delta function $\func{\symDelta}{\phi}$, considering the chosen reciprocal interpolation for the density distribution according to \myeqref{eq:rhoEff}.

\paragraph{Evaporative cooling}
In the energy equation \eqref{eq:heat_transfer}, we consider evaporative cooling, i.e., the heat absorbed during evaporative phase change, as a regularized interface flux
\begin{equation}
	\label{eq:vapor_heat_loss}
	\func{\evaporCooling}{\Bx,t} = -\func{\mDot}{T}\,\hv\,\func{\deltaRhoCp}{\phi}
\end{equation}
with the specific latent heat of evaporation $\hv$ and the parameter-scaled delta function $\deltaRhoCp$, considering the interpolation function for the volume-specific heat capacity, described in~\cite{much2023}:
\begin{equation}
	\func{\deltaRhoCp}{\phi} = \symDelta(\phi)\,\rhoCpEff(\phi) c_{\rhoCp}\quad\text{ with }c_{\rhoCp} = \frac{3}{\rhoCpG^2+\rhoCpL\rhoCpG+\rhoCpL^2}\,.
	\label{eq:deltaRhoCp}
\end{equation}

\subsection{\replaced{C}{Refined formulation of a c}onsistent level-set transport velocity \added{formulations} for a diffuse evaporation-induced velocity jump}
\label{sec:level_set_transport_velocity}

{
\color{black}
A critical modeling component of phase change across the liquid-gaseous interface $\Gamma$ is an accurate expression for the level-set transport velocity $\uGamma$, used in \myeqref{eq:transport}. This quantity ultimately governs the interface evolution and thus affects the overall dynamics. In \ref{app:level_set_transport_velocity}, we propose two new formulations for $\uGamma$, extending our previous work~\cite{schreter2024consistent}, and discuss the overall five possible variants in a theoretical study. Here, we summarize the two distinct variants used in the numerical studies in Sections~\ref{sec:numerical_examples_two_phase_flow} and \ref{sec:numerical_examples_pbfam}:

\begin{itemize}
	\item \emph{Variant 1} computes the  level-set transport velocity as
	\begin{equation}
		\uGamma^{(\text{V1})}(\Bx,t) = \Bu(\Bx,t) + \frac{\mDot(\Bx,t)}{\rhoEff(\Bx, t)}\,\nGamma(\Bx,t) \qquad \text{ for }\Bx \text{ in }\Omega\,.
		\label{eq:transport_vel_local_continuous}
	\end{equation}
	It is suitable for flat or slightly curved interfaces with a small interface thickness to curvature radius ratio. It involves the evaluation of only local quantities, which makes it attractive in terms of computational efficiency.
	\item \emph{Variant 2} considers an extension of the velocity from the liquid end of the interface region $\liqP{\Bx}$ (defined as the projection of a point $\Bx$ along the interface normal $\nGamma$ to the level-set isocontour where $H_\phi(\phi)=1$) and computes the level-set transport velocity as
	\begin{equation}
		\uGamma^{(\text{V2})}(\Bx,t) = \Bu(\liqP{\Bx}(\Bx,t) ) + \frac{\mDot(\Bx_\Gamma(\Bx,t))}{\rhoL}\,\nGamma(\Bx,t)
		\qquad
		\text{ for }\Bx \text{ in }\Omega\,.
		\label{eq:transport_vel_extension_liquid}
	\end{equation}  
To avoid a strong variation of the evaporative mass flux $\mDot$  in Eq.~\eqref{eq:transport_vel_extension_liquid} across the finite interface region, 
	we perform an extension algorithm from its value at the interface midplane, i.e., $\Bx_\Gamma(\Bx,t)$, following~\cite{much2023}. This approach ensures that the transport velocity according to variant 2 remains constant across the interface region. 
	This property benefits the level-set framework by reducing the need for frequent reinitialization steps~\cite{coquerelle2016fourth}.
\end{itemize}
Variant 1 and 2 are mathematically consistent such that the modeling error of the diffuse formulation vanishes as the interface thickness decreases ($\epsilon\rightarrow0$).  
In~\cite{schreter2024consistent}, it was concluded that variant~2 provides the best trade-off between accuracy and computational cost in 2D.

} 

\section{Numerical study of anisothermal two-phase flow with evaporation}
\label{sec:numerical_examples_two_phase_flow}
For evaluating the strengths and weaknesses of the diffuse framework for anisothermal two-phase flow with evaporation, with theoretical details provided in Section~\ref{sec:two_phase_flow} and the numerical framework in
\ref{app:numerical_framework}, we compute two benchmark examples: (i) the one-dimensional Stefan problem (Section~\ref{sec:one_dimensional_stefan_problem}) and (ii) the film boiling problem (Section~\ref{subsec:film_boiling}). For benchmark (i), an analytical solution is available that allows us to verify the numerical solution with respect to the evaporation-induced motion of the liquid--vapor interface. Thereby, the evaporation-induced temperature field and resulting motion of the liquid--vapor interface is investigated for quasi-one-dimensional conditions.
In benchmark (ii), the complexity is increased by simulating a fully coupled anisothermal two-phase flow problem with evaporation. The film boiling example involves features such as vapor formation and bubble pinch-off behavior, which are also needed in prospect of simulating the melt--vapor interaction in metal additive manufacturing. 

For calculating the evaporative mass flux, we use the model by Tanasawa~\cite{tanasawa1991advances}, representing a simplified version of the model by Schrage~\cite{schrage1953theoretical}. This model was also adopted in related publications~\cite{hardt2008evaporation,hosseini2019numerical}. \replaced{Notably, any evaporation model can be incorporated, for example, by evaluating the evaporative mass flux from the energy balance at the interface, assuming that the interface temperature corresponds to the boiling temperature~\cite{lee2017direct}.}{It is worth noting that in principle the incorporation of any other evaporation model is possible, e.g., by evaluating the evaporative mass flux from the energy balance at the interface under the assumption that the interface temperature corresponds to the boiling temperature~\cite{lee2017direct}.} \deleted{In this context,} We demonstrate in \ref{remark_1} that the Tanasawa model \replaced{serves}{can be interpreted} as a penalty method for enforcing the boiling temperature at the interface.

The Tanasawa model assumes a relation of the evaporative mass flux to the (interface) temperature $T$ and the boiling temperature $\Tv$ as follows
\begin{equation}
	\label{eq:mDotTanasawa}
	\mDot(T) = \evaporationHeatTransferCoefficient\,(T-\Tv)
\end{equation}
with the evaporation heat transfer coefficient $\alpha_v$ calculated from
\begin{equation}
	\evaporationHeatTransferCoefficient = \frac{2\evaporationCoefficient}{2-\evaporationCoefficient}\frac{\hv}{\sqrt{2\pi \specificGasConstant}}\frac{\rhoG}{\Tv^{3/2}}\,.
\end{equation}
Here, $0<\evaporationCoefficient\leq1$ is the evaporation coefficient, representing the fraction of molecules transferred from the liquid
phase to the vapor phase during evaporation, $\hv$ the latent heat of evaporation, $\rhoG$ the vapor density, $\specificGasConstant=R/M$ the specific gas constant with the molar gas constant $R = 8.314 \, \si{\joule\per\mole\per\kelvin}$ divided by the molar mass $M$ and $\Tv$ the boiling temperature.

If not stated otherwise, the following assumptions hold for the numerical simulations. 
\begin{itemize}
	\item 
	For evaluating the interface temperature to compute the evaporative mass flux, we employ an extension algorithm as proposed in~\cite{schreter2024consistent, much2023}. For a given point $\Bx$, we determine the projected point $\Bx_\Gamma$ along the interface normal direction $\nGamma$ to the interface midplane ($\phi=0$). Then, we evaluate the temperature at this point $T(\Bx_\Gamma(\Bx))$ to compute the evaporative mass flux $\mDot(\Bx_\Gamma(\Bx))$ according to \myeqref{eq:mDotTanasawa}. In~\cite{much2023} it has been demonstrated
	 that the extension algorithm yields a higher accuracy compared to local evaluation for regularized modeling of temperature-dependent interface fluxes, such as present in the terms for the evaporative dilation rate~\eqref{eq:evaporDil}, the level-set transport velocity (cf. Section~\ref{sec:level_set_transport_velocity}) or evaporative cooling~\eqref{eq:vapor_heat_loss}.
	\item 
	For computation of the level-set transport velocity, we use variant 2 according to \myeqref{eq:transport_vel_extension_liquid}.
	\item  If units are omitted in this section, they are assumed to correspond to SI standards, i.e., kg, m, s, K. 
\end{itemize}

\subsection{One-dimensional Stefan problem}
\label{sec:one_dimensional_stefan_problem}

\begin{figure}[tb!]
	\centering
	\includegraphics{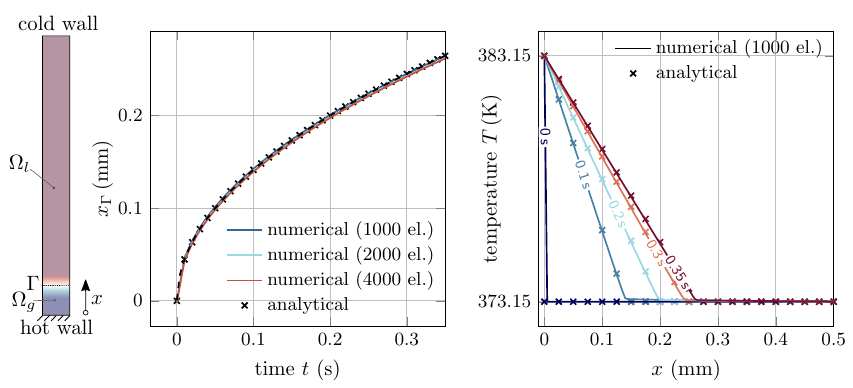}
	\caption{One-dimensional Stefan problem: (left) problem setup; (center) temporal movement of the sharp interface ($x_\Gamma=x(\phi=0)$) for different mesh resolutions; (right) temperature profiles at different stages for a mesh with 1000 finite elements. The numerical solution is verified against the analytical solution.}
	\label{fig:stefan_problem_with_heat}
\end{figure}

\def\Twall{T^{\text{(wall)}}}
In the one-dimensional Stefan problem, we study the behavior of conductive heat transfer from an isothermal hot wall through a vapor phase to a (quasi-)isothermal liquid phase. The problem setup is illustrated in the left panel of Figure~\ref{fig:stefan_problem_with_heat}. Since the temperature at the wall $\Twall$ is above the boiling point $\Tv$, liquid material evaporates, which induces a motion of the liquid--vapor interface. This benchmark has already been considered in previous contributions~\cite{hardt2008evaporation, lee2017direct}. For verification of the numerical results, an analytical solution for the problem is available, see e.g.~\cite{alexiades2018mathematical}. The interface location is calculated as
\begin{align}
	x_{\Gamma}(t) = 2\beta\sqrt{\gasP{a}\,t}
\end{align}
with the thermal diffusivity of the vapor phase
$\gasP{a}=\kG/(\rhoG\,\cpG)$. The growth constant $\beta$ is implicitly determined from 
\begin{equation}
	\beta \exp(\beta^2) \erf(\beta) - \cpG\,\frac{\Twall-\Tv}{\hv\,\sqrt{\pi}}=0\,
\end{equation}
with the error function $\erf(\beta)$. This nonlinear equation is solved numerically using a Newton-Raphson scheme.
The analytical solution for the temperature reads
\begin{equation}
	T(x,t)=\max\left(\Twall + \frac{\Tv-\Twall}{\erf(\beta)}\,\erf\left(\frac{x}{2\sqrt{\gasP{a}\,t}}\right), \Tv\right)\,.
\end{equation}
For our numerical model, we choose the parameters and problem setup according to~\cite{hardt2008evaporation}.
The one-dimensional domain is restricted to $\Omega=x\in[0,\SI{e-3}{m}]$. The wall at $x=0$
is initially covered by a vanishingly thin vapor film, realized by an initial position of the discrete interface at $x_{\Gamma}^{(0)}:=x(\phi=0, t=0)=\num{e-6}$, and is heated to $\Twall=\Tv+\Delta T$ with the boiling point $\Tv=\SI{373.15}{K}$ and a temperature increase $\Delta T=\SI{10}{K}$. The temperature at the opposite wall is fixed to  $\Tv$. The initial temperature is set to $T^{(0)}=\max(\Twall - \Delta T\,x/x_{\Gamma}^{(0)};\Tv)$. 
Homogeneous Dirichlet boundary conditions for the velocity are assumed. The fluid is initially at rest ($u^{(0)}=0$). We assume equal densities of the liquid and vapor phase ($\liqP{\rho}=\gasP{\rho}=1$) and therefore expect the fluid velocity (and pressure) to remain zero.
In order to ensure that the interface temperature is 
close to $\Tv$ with the diffuse interface model, we assume a high thermal diffusivity of the liquid phase by choosing $\kL=\SI{1}{W/(mK)}$. The remaining parameters are chosen as $\cpG=\cpL=\SI{e3}{J/(kgK)}$, $\kG=\SI{e-2}{W/(mK)}$, $\hv=\SI{e6}{J/kg}$, $M=\SI{0.018}{kg/mol}$, $\evaporationCoefficient=\num{5e-3}$, $\muL=\SI{1}{kg/(ms)}$, $\muG=\SI{1e-10}{kg/(ms)}$. Note that the accommodation coefficient $\evaporationCoefficient$ used in \myeqref{eq:mDotTanasawa} is calibrated such that the underlying assumption of the analytical solution, i.e., that the interface temperature remains at the saturated temperature $\Tv$, is approximately replicated (see explanation in \ref{remark_1}).
The simulation is performed for the time period $0\leq t \leq 0.35$\,s with a constant time step size of \num{5e-5}\,s. 
Three different uniform meshes discretized by 1000, 2000 and 4000 elements are employed. The interface thickness parameter is chosen proportional to the element edge length $h$ as $\epsilon=2\,h$.
This results in a discretization of the regularized interface fluxes in the interface region by approximately 12 elements.

In the central panel of Figure~\ref{fig:stefan_problem_with_heat} the numerically predicted temporal evolution of the interface position ($\phi=0$) is illustrated for the three different discretizations. It is in perfect agreement with the analytical solution. Similarly, also the temperature profiles at different stages, illustrated in the right panel of Figure~\ref{fig:stefan_problem_with_heat}, coincide with the analytical solution. The agreement of the results obtained using the three different meshes confirms grid convergence.

\subsection{Film boiling}
\label{subsec:film_boiling}

\def\Twall{T^{\text{(wall)}}}

The investigation of film boiling, which entails the formation of new interfaces from a stable superheated vapor layer, is a popular benchmark for validating the computation of evaporating flows~\cite{hardt2008evaporation,welch2000volume,gibou2007level}. The film boiling process encompasses challenging features from a numerical point of view such as bubble formation and pinch-off behavior. In this study, we employ both 2D and 3D simulations to showcase the capabilities of our framework and to validate it against numerical results reported by other research groups.

\begin{figure}[tb!]
	\includegraphics{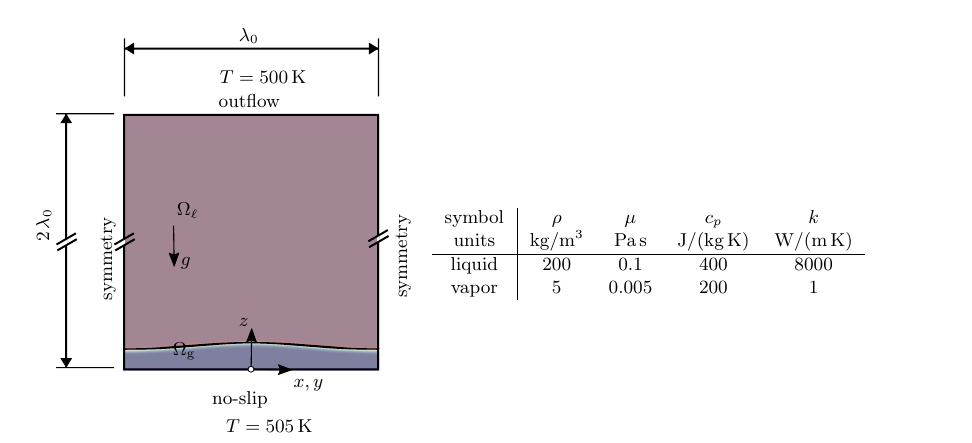}
	\caption{(left) Sketch of the film boiling example; (right) thermophysical properties for the film boiling example according to~\cite{welch2000volume,hardt2008evaporation}.}
	\label{fig:film_boiling_problem_setup}
\end{figure}

The problem setup and the thermophysical properties of the phases are summarized in Figure~\ref{fig:film_boiling_problem_setup}.   We consider the domain $\Omega=x,y \in\left[-\lambda_0/2,\lambda_0/2\right], z\in\left[0,2\,\lambda_0\right]$. 
To trigger a bubble formation through a Rayleigh-Taylor instability, we add a small perturbation of the most instable wavelength $\lambda_0=2\pi\sqrt{(3\surfTenCoeffConst)/(g\,(\rhoL-\rhoG))}=0.07868$\,m to a flat interface in the initial state. 
The vertical component of the initial interface location is specified according to~\cite{hardt2008evaporation} as
\begin{equation}
	z^{(0)}_\Gamma(\boldsymbol{x}) = z_0 + \Delta z \cos\left(\frac{2\pi r(\boldsymbol{x})}{\lambda_0}\right)
\end{equation}
with  $r(\boldsymbol{x})$ representing the shortest distance of a point $\boldsymbol{x}$ to the $z$-axis, and the vertical coordinates chosen as 
$z_0=\cfrac{4\lambda_0}{128}$ and $\Delta z= \cfrac{\lambda_0}{128}$~\cite{lee2017direct}. 
Gravitational forces with $\boldsymbol{g}=-9.81\,\eZ$ and temperature-independent surface tension with a surface tension coefficient of $\surfTenCoeffConst=\SI{0.1}{N/m}$~\cite{hardt2008evaporation} are considered. 
The fluid is initially at rest ($\velocity(\Bx,t=0)=\Bzero$).  
The initial temperature is set to $T(\Bx,t=0)=\max(\Twall - \Delta T\,z/z_{\Gamma}^{(0)};\Tv)$, 
which represents a linear function in the vapor phase and a constant function in the liquid phase. We specify the wall temperature to $\Twall=\SI{505}{K}$ and the boiling temperature to $\Tv=\SI{500}{K}$ according to~\cite{hardt2008evaporation}, resulting in a temperature difference $\Delta T=\Twall-\Tv=\SI{5}{K}$.
At the base plate ($z=0$) the temperature is fixed to $\Twall$, and no-slip conditions for the flow velocity ($\velocity(\Bx,t)=\Bzero$) are assumed. 
At the top domain boundary ($z=2\,\lambda_0$) we fix the temperature to $\Tv$, and assume a zero pressure outlet for the flow field.  
In addition, to avoid numerical difficulties for the two-phase flow framework related to potential inflow at the top domain boundary, we adopt the assumption that inflowing fluid is in a liquid state. This is strongly enforced by setting the discrete values of $\bar{\phi}=1$ for inflow parts of the domain boundary, i.e., where $\boldsymbol{u} \cdot \hat{\Bn} \leq 0$ holds.
Along the vertical domain boundaries ($x=y=\const$) we consider symmetry boundary conditions.
To ensure the interface temperature remains close to $\Tv$, we assume a high thermal diffusivity of the liquid phase by choosing $\kL=\SI{8000}{W/(mK)}$ and an accommodation coefficient of $\evaporationCoefficient=1.0$.
The molar mass is assumed as $M=\SI{0.018}{kg/mol}$. 
In addition, the latent heat of evaporation is considered as $\hv=\SI{e4}{J/kg}$~\cite{hardt2008evaporation}.
	
\begin{figure}[tb!]
	\centering
	\includegraphics{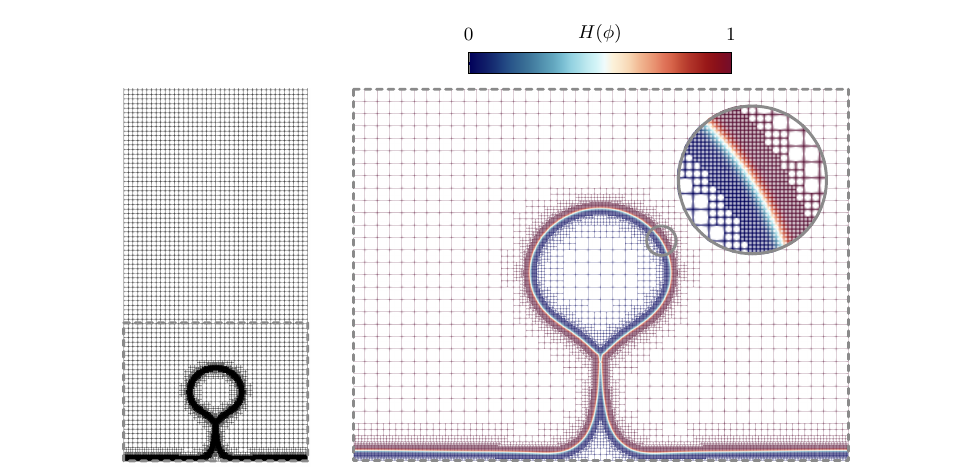}
	\caption{2D simulation of film boiling: adaptively refined finite element mesh at a representative stage. The colors in the right panel indicate the smooth heaviside function of the level set.}
	\label{fig:2D_film_boiling_mesh}
\end{figure}

\subsubsection{2D Simulation}

The 2D simulation is performed for the time period $0\leq t \leq 7$\,s with a constant time step size of \SI{e-4}{s}. 
%
For space discretization, an initially uniform mesh with $40\times 80$ finite elements is adaptively refined by four successive bisection steps in the interface region, resulting in an element edge length between $\SI{1.23e-4}{m}$ and $\SI{1.97e-3}{m}$. A snapshot from the resulting adaptively refined mesh in a representative stage of the simulation is illustrated in Figure~\ref{fig:2D_film_boiling_mesh}. The interface thickness parameter is chosen as $\epsilon=\SI{2e-4}{m}$. Considering a
refined mesh for the level-set framework by subdividing it $\nsub= 2$ times results in a resolution of the interface region by approximately 20 elements for the level-set field.

\begin{figure}[tb!]
    \includegraphics{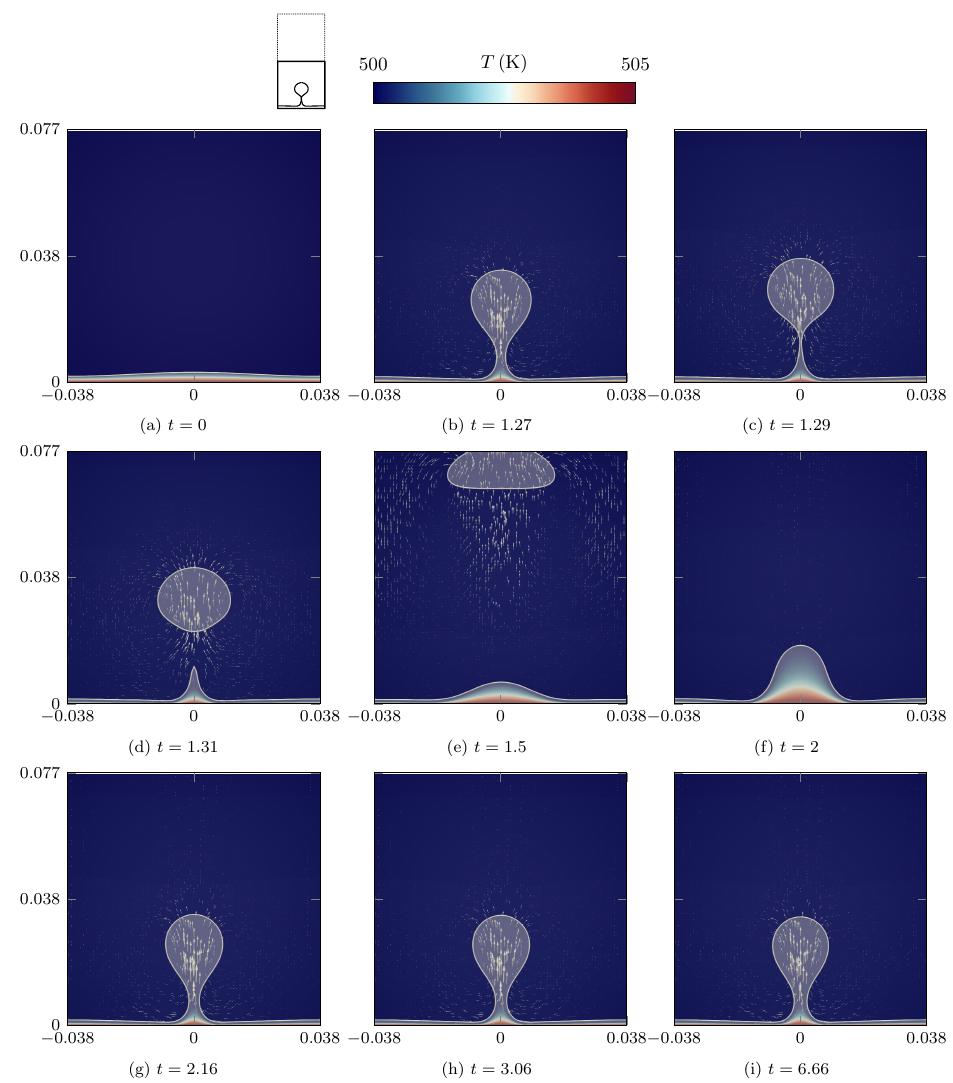}
	\caption{2D film boiling: time series showing the temperature field and the velocity field during bubble formation and pinch-off in a section of the computational domain of $\left[\lambda_0 \times \lambda_0\right]$. The vapor domain is indicated by a opaque white region, and the velocity vectors are shown.}
	\label{fig:2D_film_boiling_snapshots}
\end{figure}

In Figure~\ref{fig:2D_film_boiling_snapshots}, snapshots show a contour plot of the temperature, velocity vectors and the interface isosurface ($\phi=0$). A repeated sequence of bubble formation and pinch-off at the symmetry plane is observed. Surface tension forces lead to flattening of the vapor film after the pinch-off. Similar bubble shapes were observed in experiments, e.g.~\cite{dhir2001numerical}. It is mentioned in~\cite{lee2017direct} that 2D simulations do not realistically model actual bubble formation, since the effect of radial curvature is missing and typically elongated vapor channels are observed which makes it difficult to obtain a converged solution~\cite{gibou2007level}. Nevertheless, it represents a reasonable benchmark case to verify the numerical framework. 

\begin{figure}[tb!]
\includegraphics{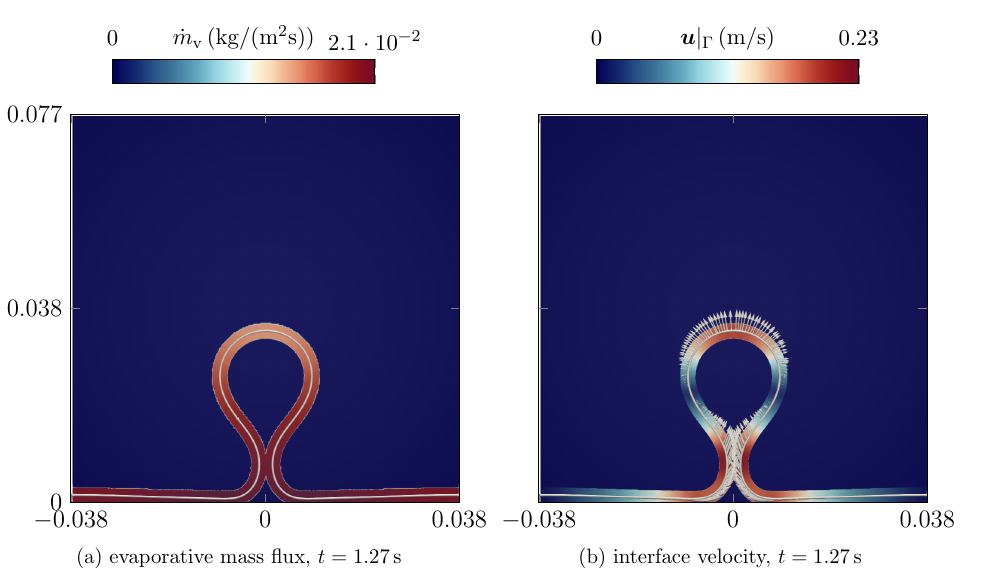}
	\caption{2D film boiling: contour plot of (left) the evaporative mass flux and (right) the level-set transport velocity at a representative stage in a section of the computational domain of $\left[\lambda_0 \times \lambda_0\right]$. In addition, the zero-level-set isosurface is shown with the corresponding vectors of the level-set transport velocity.}
	\label{fig:2D_film_boiling_evapor_flux}
\end{figure}

In the left panel of Figure~\ref{fig:2D_film_boiling_evapor_flux}, the resulting evaporative mass flux according to the Tanasawa model using the computation of the interface temperature from the extension algorithm of the zero-level-set isosurface is shown. \replaced{The}{It can be seen that the} latter is constant across the thickness within a narrow band around the interface. Similarly, in the right panel of Figure~\ref{fig:2D_film_boiling_evapor_flux}, the computed level-set transport velocity according to variant 2 is shown. Despite the complexity of the geometric shape of the interface, the velocity remains constant over the interface thickness.

\begin{figure}[tb!]
	\centering
	\includegraphics{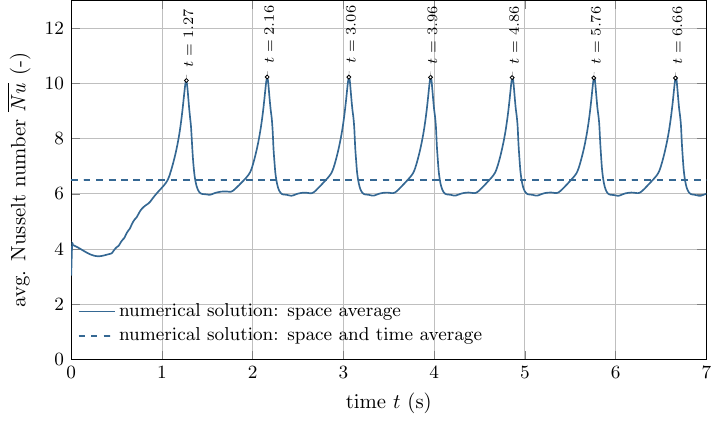}
	\caption{2D film boiling: space (and time) averaged Nusselt number evaluated at the base plate. 
		The space-time average of the numerical results \replaced{is}{can be} compared with the empirical correlation for the Nusselt number by Berenson $\Nuss^{\text{(B)}}= 0.425\,(\Gras\,\Pran/\beta_n)^{0.25}=2.62$, considering the Grashof number $\Gras=((\rhoL-\rhoG)\rhoG\,g\,\lambda_{\text{ref}}^3)/(\muG^2)=144.6$, the Prandtl number $\Pran=\cpG\,\muG/\kG=1.00098$ and the coefficient $\beta_n=\cpG \Delta T/\hv=0.1$. 
	}
	\label{fig:2D_film_boiling_nusselt}
\end{figure}

In Figure~\ref{fig:2D_film_boiling_nusselt}, we evaluate the spatially averaged Nusselt number at the base plate from the numerical simulation by
\begin{equation}
	\overline{\Nuss}=\frac{\lambda_{\text{ref}}}{\lambda_0\,\Delta T}\,\int_{-\lambda_0/2}^{\lambda_0/2} \fracPartial{T}{z}\vert_{z=0}~dx
\end{equation}
with $\Delta T = \Twall - \Tv= \SI{5}{K}$ and the reference length scale $\lambda_{\text{ref}}=\sqrt{\surfTenCoeffConst/ (g\,\left(\rhoL-\rhoG\right))}=0.00723$\,m. 
It represents a measure for the non-dimensional heat flux. A similar evaluation is reported in~\cite{lee2017direct,hosseini2019numerical}. \replaced{A}{It can be seen that a}fter the initial stage, where the vapor film builds up, a cyclic behavior between bubble pinch-offs (trough value) and bubble formation (peak values) is observed, forming a periodic pattern with a period of 0.9\,s.
In addition, we calculate the spatial-temporal average of the numerically computed Nusselt number. \replaced{This is comparable to}{It can be compared with} empirical correlation numbers, such as the one proposed by Berenson~\cite{berenson1961}. In our case, the numerically predicted spatio-temporal average of the Nusselt number ($
\overline{\Nuss}=6.5$) is significantly higher than the one obtained by the empirical Berenson correlation ($
\overline{\Nuss}=2.62$), which is attributed to the specific modeling assumption of the chosen evaporation model, as has also been observed in~\cite{hosseini2019numerical}.

\subsubsection{3D Simulation}
\label{sec:film_boiling_3D}
We simulate the film boiling example in 3D to demonstrate the versatility of our framework and its robustness in modeling complex geometrical evolutions.	The 3D simulation is performed for the time period $0\leq t \leq 5$\,s with a constant time step size of \num{2e-4}\,s. 
%
For space discretization, an initially uniform mesh with $20\times20\times40$ finite elements is adaptively refined three times in the interface region, resulting in an element edge length between $\SI{4.92e-4}{m}$ and $\SI{3.93e-3}{m}$. The interface thickness parameter is chosen as $\epsilon=\SI{5e-4}{m}$. Considering a
refined mesh for the level-set framework by subdividing it $\nsub= 2$ times, this results in a resolution of the interface region by approximately 12 elements for the level-set field. For the simulation presented in the following, the employed space discretization results in an overall maximum number of finite elements of approx. \num{670000} or \num{2.1e7} degrees of freedom in total for the level-set, temperature, pressure and velocity field. 
In contrast to the 2D simulation, we employ two additional simplifying modeling assumptions. Firstly, we use the local value of the temperature to compute the evaporative mass flux \eqref{eq:mDotTanasawa}. Secondly, the level-set transport velocity is evaluated according to variant 1 \eqref{eq:transport_vel_local_continuous}. 
While these assumptions \deleted{may} result in a less accurate solution, as discussed in~\cite{schreter2024consistent, much2023}, they are instrumental in avoiding the high computational costs in 3D associated with the extension algorithm. Our immediate focus will be on optimizing the performance of the extension algorithm to use it in large-scale 3D simulations in the future.
	
\begin{figure}[tb!]
	\includegraphics{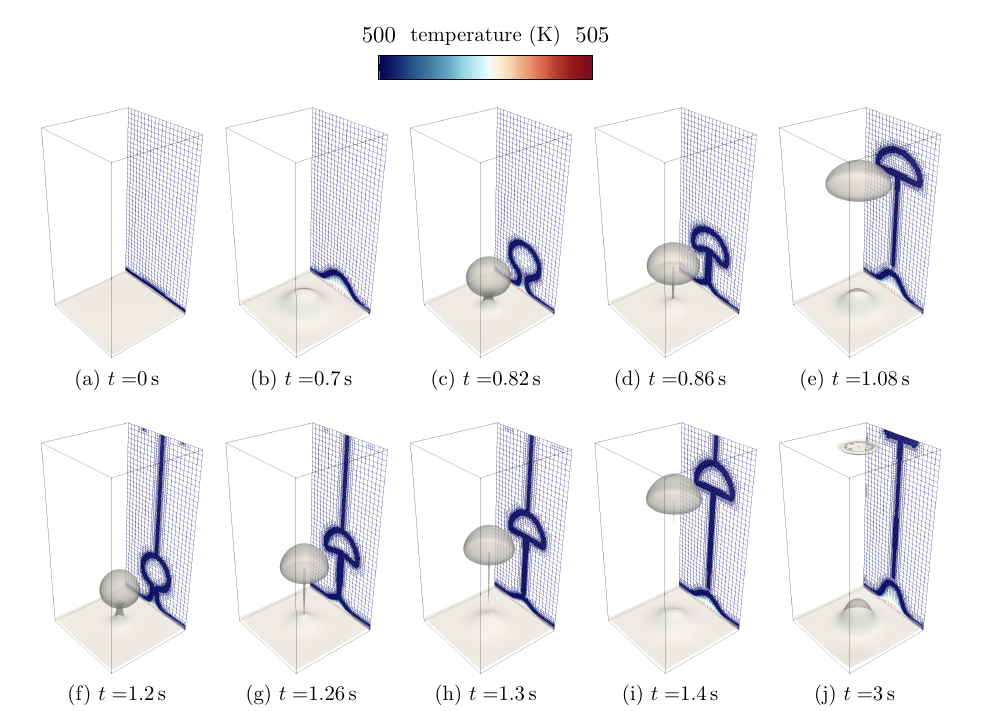}
	\caption{3D film boiling: time series showing the zero-level-set isosurface and the temperature field in a vertical section through the center during bubble formation and pinch-off.}
	\label{fig:3D_film_boiling}
\end{figure}

\begin{figure}[tb!]
		\centering
	\includegraphics{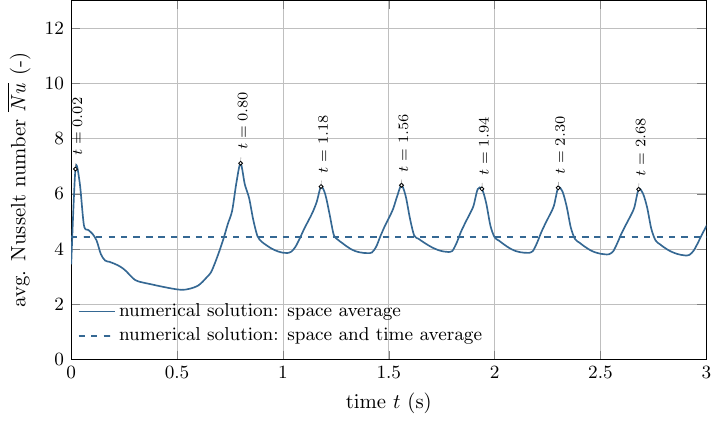}
	\caption{\added{3D film boiling: space (and time) averaged Nusselt number evaluated at the base plate. 
	}}
	\label{fig:3D_film_boiling_nusselt}
\end{figure}

In Figure~\ref{fig:3D_film_boiling}, snapshots from the 3D simulation of the film boiling example are shown. Qualitatively, the results resemble the behavior from the 2D simulations, representing a repeating sequence of bubble formation and pinch-off behavior. \added{Additionally, in Figure~\ref{fig:3D_film_boiling_nusselt}, we present the spatial and spatiotemporal average of the computed Nusselt number. Compared to the 2D simulation, the 3D simulation predicts
\begin{equation}
\overline{\Nuss}=\frac{\lambda_{\text{ref}}}{\lambda_0^2\,\Delta T}\,\int_{-\lambda_0/2}^{\lambda_0/2}\int_{-\lambda_0/2}^{\lambda_0/2} \fracPartial{T}{z}\vert_{z=0}~dx\,dy = 4.43
\end{equation}
which is closer to the empirical Berenson correlation ($\overline{\Nuss}=2.62$). 
} 

\section{A consistent multi-phase thermo-hydrodynamic melt pool model explicitly resolving the melt--vapor dynamics}
\label{sec:melt_pool}

\begin{figure}[bt!]
	\includegraphics{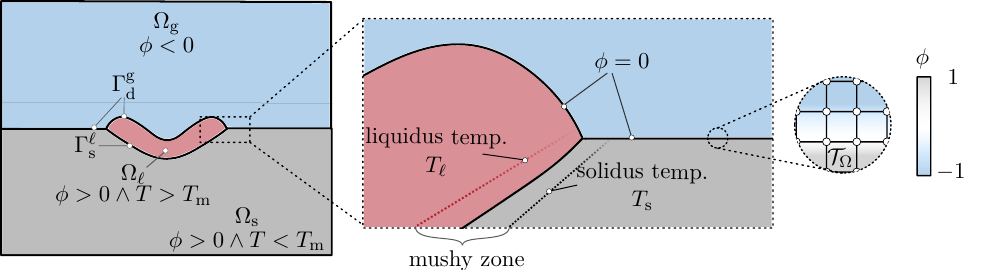}
	\caption{
		Physical domain of interest for melt pool thermo-hydrodynamics.
		The domain is decomposed into solid, liquid and gaseous phases, represented by $\OmegaS$, $\OmegaL$, $\OmegaG$, respectively. The phases are separated by the solid--liquid interface $\GammaSL$, th liquid--gaseous interface $\GammaLG$ and the solid-gaseous interface $\GammaSG$. 
		Based on the level-set function $\phi$, tracking the dense--gaseous interface $\GammaLevelSet = \GammaLG \cup \GammaSG$, and the temperature $T$, the phases are implicitly distinguished. The spatial discretization of the domain is performed by a finite element mesh~$\mathcal{T}_\Omega$.}
	\label{fig:three_phase_domain}
\end{figure}

In the following, we extend the presented anisothermal two-phase flow with evaporation framework of Section~\ref{sec:two_phase_flow} to study melt pool thermo-hydrodynamics including resolved modeling of the melt--vapor interaction. As illustrated in Figure~\ref{fig:three_phase_domain}, we assume that the Eulerian domain of interest $\Omega=\OmegaG\cup\OmegaL\cup\OmegaS\in\mathbb{R}^{n}$ with $n\in\{1,2,3\}$ is occupied by a liquid phase (melt pool) $\OmegaL$ and a gaseous (vapor) phase $\OmegaG$, both modeled as incompressible and immiscible fluids, as well as an immobile, rigid solid phase ($\OmegaS$).
Irreversible phase transition between liquid and gaseous (vapor) phase, i.e., evaporation, along the liquid-gaseous interface $\GammaLG\in\mathbb{R}^{n-1}$ as well as reversible transition between liquid and solid phase (melting/solidification) along the solid--liquid interface $\GammaSL\in\mathbb{R}^{n-1}$ \deleted{may} occur. 
We employ the level-set-based diffuse interface capturing scheme for the position of the interface between the gaseous and the solid/liquid (dense) phase, denoted as $\GammaLevelSet=\GammaSG\cup\GammaLG\in\mathbb{R}^{n-1}$. In addition, the diffuse interface between solid and liquid phase is implicitly defined by the temperature interval between solidus temperature $\Ts$ and liquidus temperature $\Tl$ of the considered material. 
The solid phase is modeled as rigid and immobile.
Therefore, to inhibit the velocities in the solid phase we consider penalization via Darcy damping forces. These model components allow us to derive a single set of equations that applies to the entire multiphase domain.

\subsection{Solid--liquid interface}
\begin{figure}[tb!]
	\centering
	\includegraphics{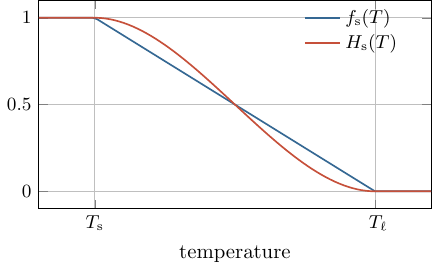}
	\caption{Illustration of the solid phase fraction \eqref{eq:solid_fraction} and the smooth Heaviside-like function  \eqref{eq:heaviside_sl}.}
	\label{fig:matprops_solid}
\end{figure}

Since the focus of this work lies on studying the melt--vapor interaction of melt pool thermo-hydrodynamics, we formulate our model based on the simplifying assumption of a rigid and immobile solid phase. 
In our model, the dense -- solid or liquid -- phase is characterized by $\phi > 0$. To distinguish between the solid and the liquid phase in the dense part of the domain, we compute the solid fraction from the temperature field as follows:
\begin{equation}
	\label{eq:solid_fraction}
	\solFrac(T(\Bx,t)) = \begin{cases}
		1 & T(\Bx,t)<\Ts\\
		\cfrac{\Tl-T(\Bx,t)}{\Tl-\Ts} & T_s\leq T(\Bx,t)\leq \Tl\,.\\
		0 & T(\Bx,t)>\Tl
	\end{cases}
\end{equation} 
Therein, the liquidus temperature $\Tl$ and the solidus temperature $\Ts$ control the width of the mushy zone, representing a finite but narrow regularized temperature interval where solid--liquid phase change (solidification/melting) takes place. From the solid fraction, we compute a smooth Heaviside-like function 
\begin{equation}
	\label{eq:heaviside_sl}
	\Hs(T)= 3\,\solFrac(T)^2-2\,\solFrac(T)^3
\end{equation}
such that $\fracTotalText{\Hs}{T}|_{\Ts}=\fracTotalText{\Hs}{T}|_{\Tl}=0$ holds to obtain a smooth interpolation of the material properties at the isotherms $T=\Tl$ and $T=\Ts$. The relations of \myeqrefs{eq:solid_fraction}{eq:heaviside_sl} are visualized in Figure~\ref{fig:matprops_solid}. Thus, the discrete interface between the liquid and the solid phase is defined as
\begin{equation}
	\GammaSL(\Bx,t) = \{\Bx \in \Omega~|~\phi(\Bx,t)\geq 0 \land \solFrac(\Bx,t)=0.5\}\,.
\end{equation}

\subsection{Effective thermophysical properties of a solid--liquid--gas system}
\label{sec:matPropsSLG}

Using \myeqref{eq:heaviside} and \myeqref{eq:heaviside_sl}\added{, we compute} volume fractions of the solid--liquid--vapor material \deleted{can be calculated} as 
\begin{equation}
	\solP{\alpha}=H_\phi(\phi)\,\Hs(T),~~\liqP{\alpha}=H_\phi(\phi)\,(1-\Hs(T))~~\text{ and }~~\gasP{\alpha}=1-\solP{\alpha}-\liqP{\alpha}=(1-H_\phi(\phi))\,.
	\label{eq:phase_fractions}
\end{equation}
In the default case, effective material properties for the three phases, such as $\muEff$, $\rhoCpEff$, $\kEff$, are evaluated as arithmetic phase-fraction weighted averages of the solid/liquid/gaseous phases based on the volume fractions~\eqref{eq:phase_fractions}
\begin{align}
	\eff{a}(\phi,T) &= 
	H_\phi(\phi)\,\left(H_s(T)\solP{a}+
		(1-H_s(T))\,\liqP{a}
		\right) + (1-H_\phi(\phi))\,\gasP{a} \\
	 &=
	\solP{\alpha}(\phi,T)\,\solP{a}+
	\liqP{\alpha}(\phi,T)\,\liqP{a} + \gasP{\alpha}(\phi,T)\,\gasP{a}\,,
	\label{eq:solid_liquid_gas_property}
\end{align}
where $\eff{a}$, $\solP{a}$, $\liqP{a}$ and $\gasP{a}$ \deleted{may} represent an arbitrary effective material property as well as properties of the solid, liquid and gaseous phase, respectively. 

For the effective density we consider the reciprocal phase interpolation function \eqref{eq:rhoEff} between the dense and the gaseous phase, for the reasons elaborated in Section~\ref{sec:matProps} and in~\cite{schreter2024consistent}, and extend it to incorporate the density of the solid phase as follows:
\begin{equation}
	\label{eq:solid_liquid_gas_density}
	\frac{1}{\eff{\rho}(\phi,T)}=\reciprocalFunc{\eff{\rho}^{(sl)}(T)}{\gasP{\rho}}=\reciprocalFunc{\Hs(T)\solP{\rho}+
		(1-\Hs(T))\,\liqP{\rho}
		}{\gasP{\rho}}\,.
\end{equation}

\subsection{Additional source terms specific to melt pool dynamics}
\label{sec:source_terms_melt_pool}

For studying melt--vapor interaction of melt pool thermo-hydrodynamics, we introduce additional right-hand side contributions in the governing equations \eqref{eq:continuity}-\eqref{eq:heat_transfer} in addition to the interface fluxes at the liquid-gas interface defined in Section~\ref{sec:interface_fluxes}. We introduce (i) a Darcy damping force $\DarcyDampingForce$ in the momentum equation \eqref{eq:momentum_balance}, designed to inhibit the motion in the solid phase, and (ii) an interface heat flux  $\laserFlux$ for modeling the illumination by a focused laser beam as typical for \pbfam{} in the energy equation \eqref{eq:heat_transfer}. We also introduce an evaporation-induced hybrid recoil force $\recoilPressureForce$, which will be discussed separately in Section~\ref{sec:hybrid_evapor}. Summarizing, the additional force contribution to the momentum balance equation is given by $\boldsymbol{f} = \surfaceTensionForce + \recoilPressureForce + \DarcyDampingForce$, while the additional fluxes to the energy equation are expressed as $s = \evaporCooling + \laserFlux$.

\begin{figure}[htbp!]
\centering
\includegraphics{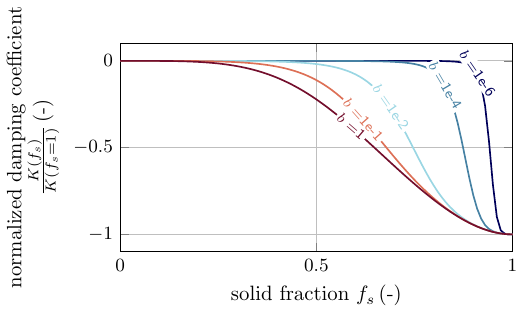}
\caption{Influence of the avoid-division-by-zero coefficient $b$ on the normalized Darcy damping coefficient $0<K(f_s)/K(1)<1$. \replaced{S}{It can be seen that s}mall values for $b$ lead to an increasingly steep gradient of the Darcy damping coefficient. 
}
\label{fig:normalized_darcy}
\end{figure}

\paragraph{Darcy damping force}
Since the focus of this contribution is on the modeling the melt--vapor interaction, the solid phase is considered as rigid and immobile by penalizing the velocity to zero. To this end, the right-hand side of the momentum balance equations \eqref{eq:momentum_balance} is augmented by a Darcy-like damping force
\begin{equation}
	\label{eq:darcy_damping_force}
\Bf_{\text{d}}=K\,\Bu \text{ with } K= -C \left(\frac{\solFrac(T)^2}{(1-\solFrac(T))^3+b}\right)\,.
\end{equation}
Within the solid phase and the solid--fluid transition region, the damping coefficient $K$ is controlled by the morphology of the mushy zone $C$ and a coefficient $b$, the latter to avoid division by zero. Too small values for $b$ \deleted{may} yield rather steep gradients in the distribution of the Darcy damping coefficient across the interface, as \replaced{visible}{can be seen} in \autoref{fig:normalized_darcy}, which is not beneficial in the context of finite element computations.  We choose  $C=\SI{e12}{kg/(m^3s)}$ and $b=1$ to obtain a sufficient constraint of the motion of the solid phase\added{, particularly relevant for the solid-gas transition region, }and a smooth transition between the mobile liquid phase and the rigid solid phase, resulting in a maximum value of the damping coefficient $K(f_s=1)=\SI{e13}{kg/(m^3s)}$. 
\added{Alternative permeability coefficients in the solid--liquid transition region could enhance the physical accuracy of mushy zone modeling, as investigated in~\cite{fadl2019numerical}. However, this aspect falls beyond the primary scope of this article.}

\paragraph{Laser heat flux}

We model the impact of the focused laser beam by a regularized interface heat flux acting on the dense-gaseous interface following a Gaussian profile
\begin{equation}
\label{eq:laser_gauss}
\laserFlux = \frac{2\,\chi_\text{L}\,P_\text{L}\,}{r_\text{L}^2 \pi}\,\langle-\nGamma\Be_{\text{L}}\rangle  \exp\left(-2\left(\frac{d_{\text{L}}(\Bx)}{r_{\text{L}}}\right)^2\right)\, \deltaRhoCp
\end{equation}
with the interface laser power absorptivity $\chi_{\text{L}}$, the laser power $P_{\text{L}}$, the laser beam radius $r_{\text{L}}$.  Macaulay brackets are denoted by $\langle\bullet\rangle$. The distance between the point $\Bx$ and the laser beam center line defined by the focus point $\Bx_{\text{L}}$ as well as the laser direction given by the unit vector $\Be_{\text{L}}$ is represented by $d_{\text{L}}(\Bx)$. The parameter-scaled delta function $\deltaRhoCp$ is computed according to \myeqref{eq:deltaRhoCp}.

\subsection{A resolved evaporation model to study melt--vapor interaction in an incompressible multiphase flow framework}
\label{sec:hybrid_evapor}

As indicated in Section~\ref{sec:intro}, the majority of existing melt pool models simplify the evaporation dynamics by solely considering the evaporation-induced recoil pressure on the melt pool surface but not spatially resolving the resulting vapor flow. To accurately capture this critical effect and gain deeper physical insights, we introduce a resolved evaporation modeling approach that incorporates vapor flow dynamics consistently within an incompressible multi-phase flow framework.
%
To this end, we incorporate the following features:

\begin{itemize}
	\item 
	The evaporative mass flux is computed according to the relation by Knight~\cite{knight1979theoretical}, i.e., 
	\begin{equation}
		\mDotPSat(T) = 0.82\,c_\text{s}\,\pSat(T)\sqrt{\frac{M}{2\,\pi\,R\,T}}
		\label{eq:mDotPSat}
	\end{equation}
	with the sticking coefficient chosen as $c_\text{s}=1$, the saturated vapor pressure $\pSat(T)$, the molar mass $M$ and the universal gas constant $R$.
	\item Despite assuming incompressibility, we account for the evaporative volume expansion~\eqref{eq:evaporDil} in the continuity equation \eqref{eq:continuity}, inducing a flow in the vapor phase and automatically resulting in a consistent recoil pressure at the liquid--vapor interface:
	\begin{equation}
		\pRecoilRegularizedMDot(\Bx,t) = \underbrace{\func{\mDotPSat}{\func{T}{\Bx,t}}^2\left(\frac{1}{\rhoG}- \frac{1}{\rhoL}\right)}_{\pRecoilMDot}\,\func{\symDelta}{\phi(\Bx,t)}\,.
		\label{eq:recoilPressureEvaporDil}
	\end{equation}
	It is emphasized that no additional terms are required to capture this effect. Instead, \myeqref{eq:recoilPressureEvaporDil} \deleted{can be shown to} follow\added{s} directly from the mass and momentum balance in \myeqrefs{eq:continuity}{eq:momentum_balance} by formulating momentum balance in interface normal direction.
	It is influenced by the employed evaporative mass flux $\mDotPSat(T)$ and the vapor density $\rhoG$ assuming that $\rhoG\ll\rhoL$ holds. 
	This effect is already captured by the anisothermal two-phase flow with evaporation framework presented in Section~\ref{sec:two_phase_flow}. 
	\item 
	Compared to relations from gas dynamics in the Knudsen layer~\cite{knight1979theoretical}  (see \ref{app:recoil_pressure}),  in line with the incompressibility assumption of our framework, we assume a temperature- and pressure-independent density of the vapor phase. Consequently, if we overestimate the vapor density, our model predicts a lower pressure jump resulting from evaporative volume expansion \eqref{eq:recoilPressureEvaporDil} compared to the model by Anisimov et al.~\cite{anisimov1995instabilities}.
	To address this discrepancy, we additionally impose an interface pressure jump to the inherently considered pressure jump via \myeqref{eq:recoilPressureEvaporDil}, denoted as \emph{hybrid recoil pressure}.

The phenomenological model by Anisimov et al.~\cite{anisimov1995instabilities}, which is typically employed in existing melt pool models~\cite{meier2021physics}, predicts an evaporation-induced recoil pressure at the melt pool surface of 
\begin{equation}
\label{eq:anisimov}
	\pRecoilAnisimov(T) = 0.54\ \pAmbient\,\exp\left(-\frac{\hv\, M}{R}\left(\frac{1}{T}-\frac{1}{\Tv}\right)\right) \text{ on }\GammaLG
\end{equation}
with the atmospheric pressure $\pAmbient$. 
Formulating \myeqref{eq:anisimov} as a regularized interface flux reads
\begin{equation}
	\label{eq:anisimov_regularized}
	\pRecoilAnisimovRegularized(\Bx,t) = \pRecoilAnisimov(T(\Bx,t))\func{\deltaRho}{\phi(\Bx,t)}\,.
\end{equation}

In our model, we assume that the overall evaporation-induced pressure jump at the melt surface should correspond to \myeqref{eq:anisimov}. Thus, we propose the definition of the hybrid recoil pressure force as
\begin{equation}
\label{eq:hybrid_recoil_pressure}
\recoilPressureForce = 
\left(\pRecoilAnisimovRegularized - \pRecoilRegularizedMDot\right)\,\nGamma\,,
\end{equation}
which is obtained by subtracting \myeqref{eq:recoilPressureEvaporDil} from \myeqref
{eq:anisimov_regularized} and multiplication with the interface normal vector $\nGamma$.
It is an additionally imposed force at the liquid--vapor interface in the momentum balance equation \eqref{eq:momentum_balance} and ensures that the resulting pressure jump aligns with Anisimov's model.
\end{itemize}

\section{Numerical study of melt--vapor interaction in \pbfam{} during stationary laser illumination of a bare metal plate}
\label{sec:numerical_examples_pbfam}

\begin{figure}[tb!]
	\centering
	\includegraphics{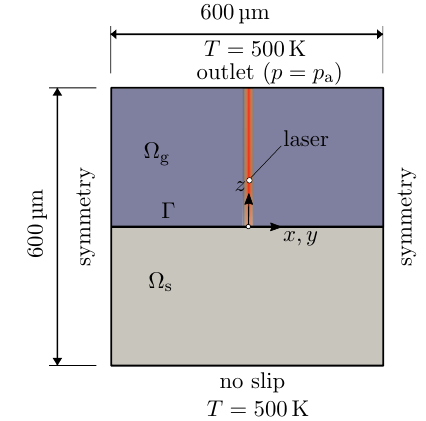}
	\caption{Sketch of the initial simulation setup for stationary laser illumination of a bare metal plate.}
	\label{fig:sketch_cunningham}
\end{figure}

As a benchmark example for the presented numerical framework in Section~\ref{sec:melt_pool} applied to \pbfam{}, we compute the stationary laser illumination of a bare \tiSixFour{} plate, inspired by the experimental setup by Cunningham et al.~\cite{cunningham2019keyhole}.
A similar setup was investigated in~\cite{much2023,meier2021physics} but without resolved modeling of the evaporation-induced flow field.
 A schematic sketch is shown in Figure~\ref{fig:sketch_cunningham}. The problem setup considers the domain $\Omega = x,y,z \in [-300\,\si{\micro\meter},300\,\si{\micro\meter}]$.
Typical material parameter values representing \tiSixFour{} are employed, according to Table~\ref{tab:param_ti64}.
The lower half of the domain ($z\leq0$) is initially covered with a solid phase, the remaining domain with a gas phase,
characterized by the signed distance function of the initial interface of a plane, i.e., $\signedDistance(\boldsymbol{x}) = -z$.
The phases are initially at rest ($\Bu^{(0)}=\boldsymbol{0}, T^{(0)}=\SI{500}{K}$). 
At the bottom domain boundary, no-slip conditions for the Navier--Stokes equations and a fixed temperature of $T= T^{(0)}$ are assumed, while on the top boundary outlet conditions at ambient pressure $p=\pAmbient$ and fixed temperature of $T= T^{(0)}$ are considered. Along the vertical domain boundaries ($x=y=\const$), symmetry boundary conditions for the flow field and adiabatic temperature conditions are assumed. 
The metal surface is exposed to a spatially fixed laser heat source ($P_\text{L}=\SI{78}{W}$, $\chi=0.35$, $r_\text{L}=\SI{70}{\micro\meter}$, $\Bx_\text{L}=\boldsymbol{0}$, $\Be_{\text{L}}\widehat{=}$ negative $z$-direction). A resolved vapor phase is considered according to Section~\ref{sec:hybrid_evapor}. In addition, we consider temperature-dependent surface tension forces in the form of Marangoni convection according to \myeqref{eq:surface_tension}.

If not stated otherwise, the following assumptions -- similar to those of Section~\ref{sec:numerical_examples_two_phase_flow} -- are employed for the numerical models: For evaluating the interface temperature to compute the evaporative mass flux \eqref{eq:mDotPSat} and the hybrid recoil pressure \eqref{eq:hybrid_recoil_pressure}, we employ an extension algorithm for the temperature value at the zero-level-set isosurface. Furthermore, for computing the level-set transport velocity, we use variant 2 \eqref{eq:transport_vel_extension_liquid}. 
We gradually increase the complexity of the simulation by studying the problem in 1D, 2D and 3D. We leverage the lower-dimensional models to perform more detailed and critical evaluations of the proposed model.

\begin{table}[tb!]
	\caption{Parameters representing \tiSixFour.}
	\label{tab:param_ti64}
	\centering
	\addtolength{\tabcolsep}{-0.2em}
	\begin{tabular}{clccc}
		\toprule
		& \emph{Parameter} & \emph{Symbol} & \emph{Value} & \emph{Unit} \\ \midrule\midrule
		&  molar mass & $M$ & \num{4.78e-2} & \si{kg / mol} \\ \midrule
		\parbox[t]{4mm}{\multirow{3}{*}{\rotatebox[origin=c]{90}{\small solid/liquid}}} & density & $\rhoL=\rhoS$ & 4087 & \si{kg/m^3} \\
		& dynamic viscosity                  & $\muL=\muS$ & \num{3.5e-3} & \si{kg/(m\,s)}\\
		& heat capacity at constant pressure & $\cpL=\cpS$ & \num{1130}   & \si{J/(kg\,K)} \\
		& thermal conductivity & $\kL=\kS$ & 28.63 & \si{W/(m\,K)} \\
		\midrule
		\parbox[t]{4mm}{\multirow{3}{*}{\rotatebox[origin=c]{90}{\small vapor/gas}}} & density & $\rhoG$ & 4.087 & $\si{kg/m^3}$\\
		& dynamic viscosity & $\muG$ & \num{3.5e-5} & \si{kg/(m\,s)} \\
		& heat capacity at constant pressure & $\cpG$ & \num{11.3} & \si{J/(kg\,K)} \\
		& thermal conductivity & $\kG$ & 0.02863 & \si{W/(m\,K)}\\
		\midrule
		\parbox[t]{4mm}{\multirow{3}{*}{\rotatebox[origin=c]{90}{\small surf.\,ten.}}}
		&  surface tension coefficient at reference temperature & $\surfTenCoeffConst$ & \num{1.493} & \si{N/m}\\
		& reference temperature & $\TRefSurfTen$ & \num{1928} & \si{K} \\
		& surface tension gradient coefficient & $\surfTenCoeffTemp$ & \num{5.5e-4} &  \si{N/(m K)} \\
		\midrule
		\parbox[t]{4mm}{\multirow{3}{*}{\rotatebox[origin=c]{90}{\small evapor.}}}& boiling point & $\Tv$ & 3133 & \si{K} \\
		& latent heat of evaporation & $\hv$ & \num{8.84e6} & \si{J / kg} \\
		& sticking constant & $c_s$ & 1 & \si{-} \\
		\midrule
		\parbox[t]{4mm}{\multirow{4}{*}{\rotatebox[origin=c]{90}{\small melt./solid.}}}
		& solidus temperature & $\Ts$ & 1933 & \si{K}  \\
		& liquidus temperature & $\Tl$ & 2200 & \si{K} \\
		&  mushy zone morphology & $K$ & \num{1e12} & \si{kg/(m^3s)} \\
		&  avoid-division-by-zero parameter & $b$ & \num{1} & \si{-} \\
	\end{tabular}
\end{table}

\subsection{1D simulation}
For the one-dimensional setup, we consider the domain along the vertical $z$-axis, i.e., $x=y=0$. This simplified setup is particular suitable for detailed verification of the numerical solution.
The 1D simulation is performed for the time period $0\leq t \leq \SI{1}{ms}$ with a constant time step size of $\SI{0.1}{\micro s}$. We consider a uniform mesh consisting of 2000 finite elements resulting in an element edge length of $\SI{0.3}{\micro m}$. The interface thickness parameter is chosen as $\epsilon=\SI{1}{\micro m}$. Considering a refined mesh for the level-set framework by subdividing it $\nsub=2$ times, the interface region is resolved by 40 elements for the level-set field.

\begin{figure}[tb!]
	\includegraphics{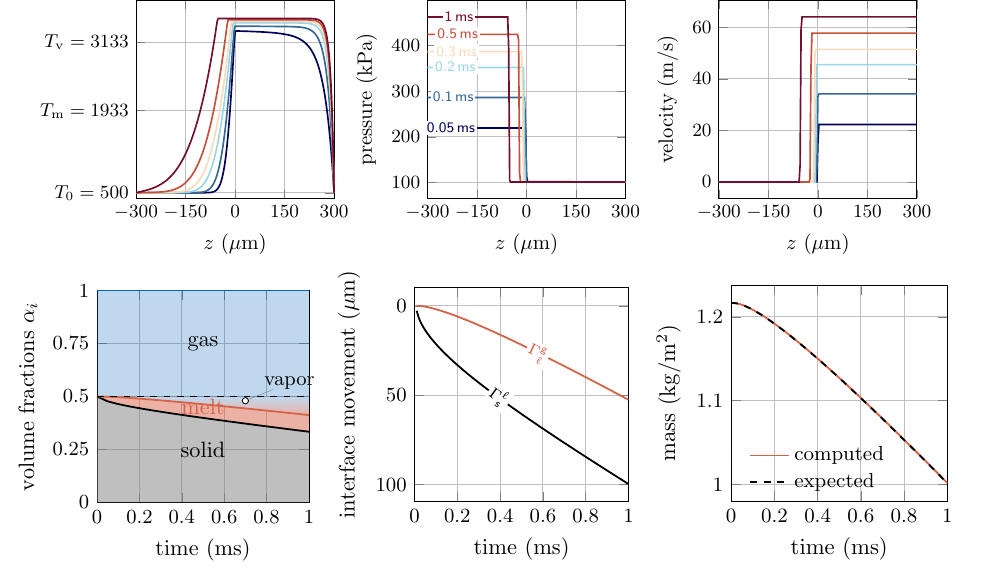}
	\caption{1D simulation of melt--vapor interaction during \pbfam{}: (top) profiles of the temperature, pressure and velocity for different time stages; 
	(bottom left) evolution of the solid, liquid and gas/vapor volume fractions \eqref{eq:phase_fractions} over time; (bottom center) temporal movement of the liquid-gas ($\GammaLG$) and the solid--liquid ($\GammaSL$) interfaces; (bottom right) evolution of the total mass over time: we compute the mass from the space integral of the current effective density $m_{\text{comp.}}(\bar{t})=\int_0^{\bar{t}}\int_\Omega\rhoEff(x,t)\,dx\,dt$ and compare it with the expected total mass considering the initial total mass $m^{(0)}$ and subtracting the mass loss due to liquid--vapor phase transition, i.e., \\
	 $m_{\text{expect.}}(\bar{t})=\underbrace{\int_\Omega\rhoEff(x,t=0)\,dx}_{m^{(0)}} - (\rhoL-\rhoG)/\rhoL\int_0^{\bar{t}} \mDot(t)\,dt$\,.
	}
	\label{fig:mp_1D}
\end{figure}

In the top panel of Figure~\ref{fig:mp_1D}, profiles of the temperature, the pressure and the velocity are shown at different stages of the simulation. Consistent with our expectations, the pressure in the dense phase as well as the velocity in the gas phase increase with rising temperature at the liquid--vapor interface. 

In the bottom left panel of Figure~\ref{fig:mp_1D} we analyze the evolution of the volume fractions \eqref{eq:phase_fractions}. \replaced{T}{Here, it can be seen that t}he volume fraction of the gaseous phase increases due to evaporative phase transition, while simultaneously the volume of the condensed metal phases decreases. 

In the bottom central panel of Figure~\ref{fig:mp_1D}, the movement of the liquid--solid interface $\GammaSL$ and the liquid-gaseous interface $\GammaLG$ is shown. Notably, the melt front propagates faster than the liquid--vapor interface. 

In the bottom right panel of Figure~\ref{fig:mp_1D}, we evaluate the temporal evolution of the total mass in the domain and compare it with the computed expected total mass, considering the initial mass and the mass loss due to liquid--vapor phase transition (which equals the mass flux leaving the top boundary of the computational domain in case of the considered incompressible model). The overlap between these two curves verifies that the interface movement predicted by the level-set framework\replaced{,}{ and in} particular\added{ly} the computed level-set transport velocity\added{,} is \replaced{accurate}{plausible}. \replaced{I}{The latter would not be the case, i}f the level-set transport velocity did not correctly take into account the evaporative mass flux\replaced{, the curves would not align}{.} 
This demonstrates that our framework is mass conserving.

\subsection{2D simulation}
\label{sec:melt_pool_2d}

\begin{figure}[tbh!]
    \includegraphics{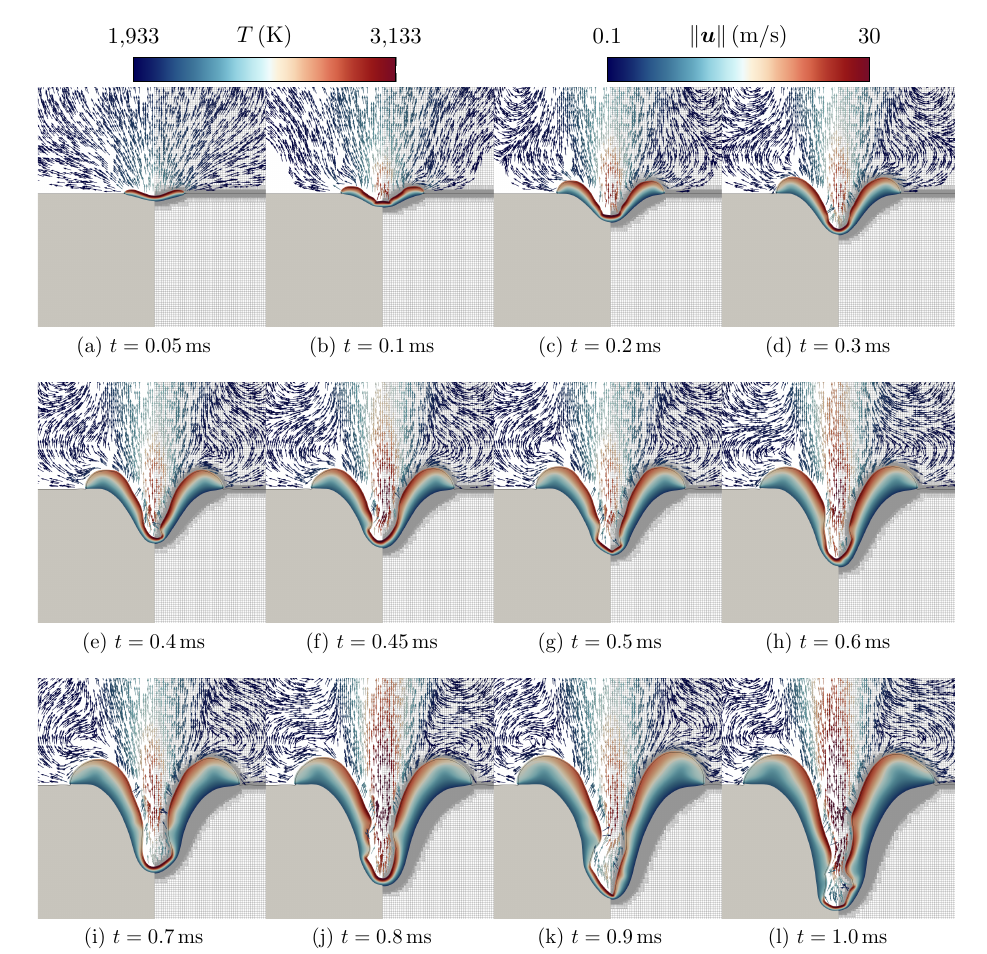}
	\caption{2D simulation of melt--vapor interaction during \pbfam{}, using level-set transport velocity \emph{variant 2}~\eqref{eq:transport_vel_extension_liquid}: time series showing the temperature in the melt pool and the vapor velocity field (vapor jet) in a section of the computational domain. \replaced{O}{It can be seen that o}scillations in the melt pool morphology are reflected in the vapor jet.}
	\label{fig:2D_melt_pool_vapor_variant2}
\end{figure}

\begin{figure}[tbh!]
	\includegraphics{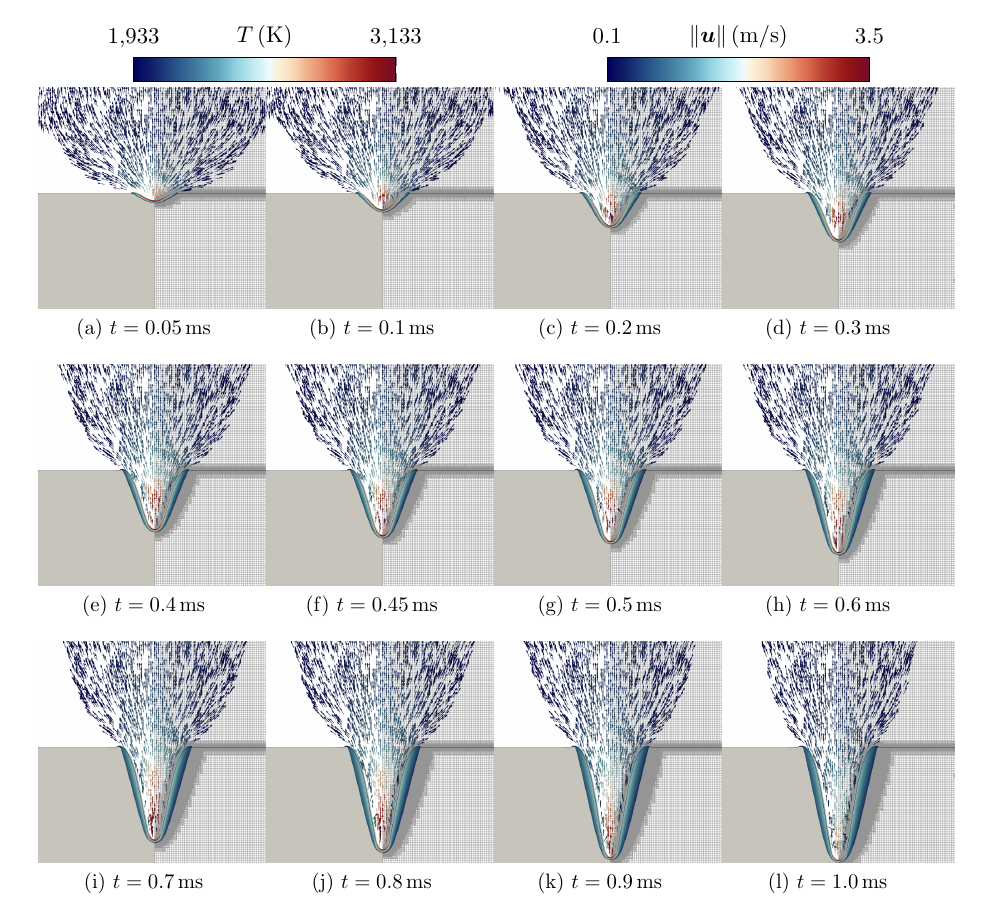}
	\caption{2D simulation of melt--vapor interaction during \pbfam{}, using level-set transport velocity \emph{variant 1}~\eqref{eq:transport_vel_local_continuous}: time series showing the temperature in the melt pool and the vapor velocity field (vapor jet) in a section of the computational domain. \replaced{T}{It can be seen that t}he melt pool  remains stable and unrealistically thin.}
	\label{fig:2D_melt_pool_vapor_variant1}
\end{figure}

\begin{figure}[tb!]
    \includegraphics{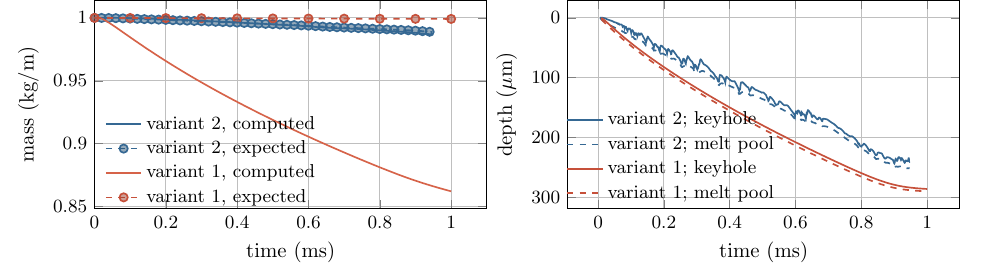}
	\caption{2D simulation of melt--vapor interaction during \pbfam{}, comparison between level-set transport velocity \emph{variant 2}~\eqref{eq:transport_vel_extension_liquid} and \emph{variant 1}~\eqref{eq:transport_vel_local_continuous}: (left) temporal evolution of the total mass. The meaning of \emph{computed} and \emph{expected} is explained in the caption of Figure~\ref{fig:mp_1D}; (right) Temporal evolution of the melt pool morphology (keyhole and melt pool depth).}
	\label{fig:2D_melt_pool_evapor_mass}
\end{figure}
For the two-dimensional setup, we consider the $x$-$z$ section (cf.~Figure~\ref{fig:sketch_cunningham}). The 2D simulation is performed for the time period $0\leq t \leq \SI{1}{ms}$ with a constant time step size of $\SI{0.01}{\micro s}$. 
For space discretization, an initially uniform mesh with $64\times 64$ finite elements is adaptively refined 
three times in the interface region, resulting in an element edge length between $\SI{0.58}{\micro m}$ and $\SI{9.37}{\micro m}$. 
The interface thickness parameter is chosen as $\epsilon=\SI{1}{\micro m}$. 
This results in a resolution of the interface region by approximately 10 elements for the level-set field. \added{The computational effort is summarized in \ref{app:comp_effort}.}

Figure~\ref{fig:2D_melt_pool_vapor_variant2} presents snapshots from the simulation, depicting the temperature distribution in the molten metal and the velocity vectors of the vapor jet at different time steps.
 Typical characteristics for \pbfam{} processing in the keyhole mode \replaced{are}{can be} observed: Initially, upon attaining the boiling temperature, the vapor depression gradually forms and transitions into a deep and narrow keyhole. During this stage, the vapor flow field remains stable. However, once the molten metal starts to fluctuate -- between recoil pressure, surface tension and geometry-dependent laser absorption -- transitioning to an unstable keyhole the fluctuating melt behavior is also reflected in the vapor flow. 
\deleted{Concluding, our results indicate a strong coupling between the dynamics of vapor and molten metal flow, as also observed in experimental results by Bitharas et al.~\cite{bitharas2022interplay}.}

\added{A quantitative comparison of Figure~\ref{fig:2D_melt_pool_vapor_variant2} with experimental results by~\cite{cunningham2019keyhole} is currently challenging due to certain physical effects that have not yet been incorporated into the model. Specifically, the laser energy input, as defined by Eq.~\eqref{eq:laser_gauss}, does not yet account for attenuation, absorption, or scattering effects caused by the evolving vapor plume~\cite{greses2004plume}. This effect becomes particularly relevant in the keyhole regime. A more advanced approach, such as ray-tracing as presented in~\cite{flint2024version}, could capture these interactions but introduces additional computational effort.
Additionally, the current model does not yet include the influence of shielding gas inlet velocity, which can impact melt--vapor interaction as shown in~\cite{reijonen2020effect}. Both aspects, the interaction of laser energy with the vapor plume and the effect of shielding gas flow, are recognized as important refinements and will be addressed in future research. Despite these pending extensions, the present model already captures the
strong coupling between the dynamics of vapor and molten metal flow, as also observed in experimental results by Bitharas et al.~\cite{bitharas2022interplay} with a mathematically consistent approach, providing a solid foundation for further development.
}

We emphasize the critical impact of the level-set transport velocity choice by conducting the same simulation using the level-set transport velocity variant 1 \eqref{eq:transport_vel_local_continuous}. This formulation is known from~\cite{schreter2024consistent} to be suitable only for flat or slightly curved surfaces with finite values of interface thickness. Snapshots from the simulation are illustrated in Figure~\ref{fig:2D_melt_pool_vapor_variant1}. Due to the violation of the assumption of low curvature in melt pool dynamics given the complex melt pool morphology, the total evaporated mass is overestimated.
This discrepancy is evident when analyzing the evolution of mass, displayed in the left panel of Figure~\ref{fig:2D_melt_pool_evapor_mass}, where a notable difference is observed between the computed and expected value -- an issue that does not arise when employing level-set transport velocity variant 2.

Consequently, a significantly different behavior \replaced{is obtained}{can be observed} compared to the results of Figure~\ref{fig:2D_melt_pool_vapor_variant2}. 
Specifically, according to Figure~\ref{fig:2D_melt_pool_vapor_variant1}, the total evaporated mass is overestimated and the melt phase appears unrealistically thin (e.g. by comparing Figure~\ref{fig:2D_melt_pool_vapor_variant2}l with \ref{fig:2D_melt_pool_vapor_variant1}l). In addition, the vapor flow velocity is extremely low and the melt pool surface remains stable without any fluctuations even under keyholing conditions. This stable behavior is further illustrated by the evolution of the melt pool and keyhole depths shown in the right panel of Figure~\ref{fig:2D_melt_pool_evapor_mass}.

\subsection{3D simulation}
\label{sec:melt_pool_3d}

The 3D simulation is performed for the time period $0\leq t \leq \SI{1}{ms}$ with a constant time step size of $\SI{0.01}{\micro s}$. 
For space discretization, an initially uniform mesh with $32\times 32 \times 32$ finite elements is adaptively refined 
three times in the interface region, resulting in an element edge length between $\SI{2.3}{\micro m}$ and $\SI{18.8}{\micro m}$. 
The interface thickness parameter is chosen as $\epsilon=\SI{2}{\micro m}$. 
This results in a resolution of the interface region by approximately 7 elements for the level-set field.  

\added{The computational effort is summarized in \ref{app:comp_effort}.}
In order to manage computational costs for 3D simulations effectively, we have chosen not to employ the extension algorithm  for computing temperature-dependent fluxes or the level-set transport velocity for the reasons described in Section~\ref{sec:film_boiling_3D}. \added{Therefore, the local value of the temperature is used to compute the evaporative mass flux \eqref{eq:mDotPSat} or the hybrid recoil pressure force~\eqref{eq:hybrid_recoil_pressure}, and the level-set transport velocity is evaluated according to variant 1 \eqref{eq:transport_vel_local_continuous}. }

\begin{figure}[tbp!]
	\includegraphics{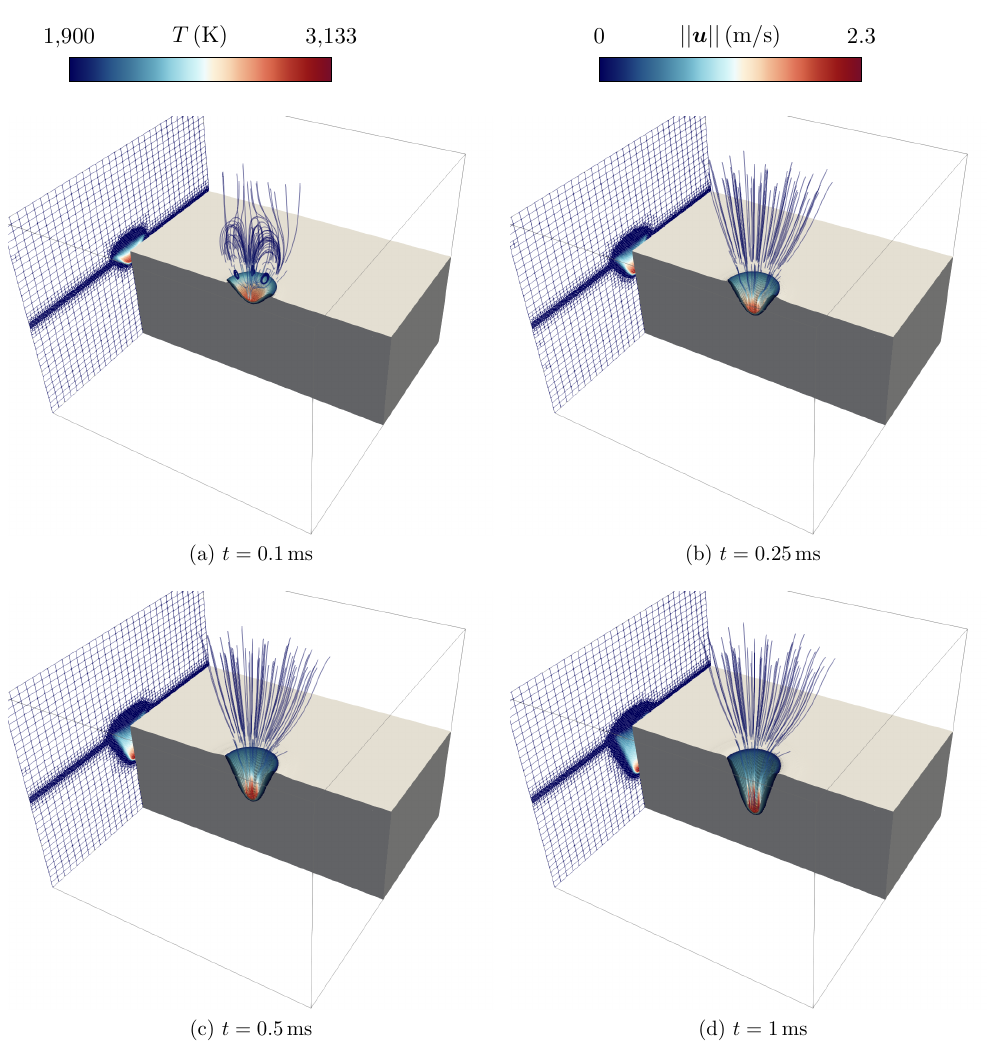}
\caption{3D simulation of stationary laser illumination, using level-set transport velocity \emph{variant 1} according to \myeqref{eq:transport_vel_local_continuous}: time series illustrating the temperature in the melt pool, streamlines of the flow field and the finite element mesh.}
\label{fig:3D_melt_pool_vapor_variant1}
\end{figure}

In Figure~\ref{fig:3D_melt_pool_vapor_variant1}, snapshots from the 3D simulation are presented, indicating melt pool formation and vapor flow as a result of resolved modeling of evaporative phase change. However, attributed to the simplifying assumptions, the resulting flow velocities are unrealistically low and the melt pool shape remains steady without exhibiting typical fluctuations throughout the simulation. As demonstrated in the 2D simulation,  \deleted{it is expected that }these artifacts \replaced{are expected to}{will} disappear\added{,} and a consistent melt pool behavior \replaced{will}{can} be predicted once a more efficient implementation of the extension algorithm is available\replaced{. This will enable the use of }{, allowing }the most accurate variant \deleted{to be employed} for evaluation of the level-set transport velocity according to variant 2 \eqref{eq:transport_vel_extension_liquid}.

\section{Conclusion}
\label{sec:conclusion}

We have developed a high-fidelity mesoscale multi-phase melt pool model tailored to the study of the melt--vapor dynamics in metal laser powder bed fusion (\pbfam{}) \added{or similar} processes. 
Unlike existing computational models in this field that often oversimplify evaporation dynamics by focusing solely on the pressure jump induced by evaporation in molten metal, our model explicitly resolves the formation of vapor jets. This is achieved by accounting for evaporation-induced mass flux,  volume expansion as well as resulting velocity changes across the liquid--vapor interface, and convective heat transfer within the gas phase. The numerical discretization and high-performance solution approach utilizes a matrix-free adaptive finite element framework based on the open-source finite element library \texttt{deal.II}~\cite{africa2024deal}.

To achieve this overall goal, first, we have extended our previously proposed isothermal two-phase flow framework with evaporation~\cite{schreter2024consistent} to include anisothermal conditions. We validated this extension through benchmark examples of anisothermal two-phase flow with evaporative phase change, including simulations of film boiling. Specifically, our model incorporates an anisothermal, incompressible Navier--Stokes solver coupled with a diffuse representation of the liquid--vapor interface using the well-established conservative level-set formulation by Olsson et al.~\cite{olsson2007conservative}. To ensure accurate physical solutions despite extreme density, pressure and velocity gradients across the finite liquid--vapor interface for rapid evaporation scenarios, we have incorporated new mathematically consistent formulations for regularized temperature-dependent interface source terms and the level-set transport velocity, as developed in our previous research works~\cite{schreter2024consistent,much2023}. 
These enhanced formulations involve projection algorithms of solution fields (e.g., temperature or velocity) from the sharp liquid--vapor interface onto the computational domain, significantly improving the accuracy of the diffuse framework.

Next, we have enhanced the anisothermal evaporating two-phase flow model to study coupled melt--vapor interaction in \pbfam{} processes. We have introduced a resolved evaporation formulation suitable for an incompressible multi-phase flow framework. 
It includes the consideration of the evaporative dilation rate in the continuity equation, resulting in an inherent pressure jump, and additionally introduces a hybrid recoil pressure force contribution in the momentum equation to result in an overall recoil pressure consistent with gas dynamics relations in the Knudsen layer~\cite{knight1979theoretical,anisimov1995instabilities}. In this context, a derivation of the well-known recoil pressure model according to Anisimov~\cite{anisimov1995instabilities} is shown and extended to yield the new hybrid model. 
Simulations of stationary laser illumination of a bare metal plate, resembling the experimental setup by Cunningham et al.~\cite{cunningham2019keyhole}, have demonstrated that the dynamic fluctuations of the melt pool due to recoil pressure and thermo-capillary forces \replaced{are}{can be} captured by the hybrid model despite the (from a numerical point of view favorable) incompressibility assumption of the vapor phase. 
Most notably, the use of projection algorithms to compute an accurate level-set transport velocity and regularized temperature-dependent fluxes is vital for the diffuse framework to capture the highly dynamic interaction between the melt pool and the vapor jet in a realistic and mass conserving manner.

Although our 2D simulations show promising results, the computational costs of the projection algorithm only allowed for first proof-of-concept simulations of the 3D problem so far.
We are currently focused on optimizing performance to improve the accuracy of 3D simulations in future work.
\added{Once these improvements are achieved,
	additional key physical effects  will be integrated to fully leverage the model potential of the model in studying melt--vapor interaction. These include plume-induced laser attenuation affecting energy deposition and temperature-dependent material properties. 
	Additionally, the current hybrid recoil pressure model, assuming a constant Mach number of 1 in the vapor regime, cannot capture pressure increase in deep keyholes caused by vapor confinement, which reduces the net evaporation rate. 
	Refining the model to include these effects or, alternatively, adopting a compressible flow formulation could address this limitation. 
}
Moreover, while the solid phase was assumed to be rigid and immobile, emphasizing the focus of the present study on melt--vapor interaction, \added{extending} our model \deleted{could be extended} to account for mobile powder particles, as initiated in~\cite{fuchs2022versatile}\added{, is a promising direction for future work}. The consideration of this mechanism \deleted{can} help\added{s} to explore how vapor jets affect the redistribution of powder particles in future research. \added{
	These refinements will advance the development of a unified framework, significantly enhancing the predictive accuracy of mesoscale models in \pbfam{} and bridging the gap between simulations and experimental observations~\cite{khairallah2016laser, zhu2021mixed}.
	}

\section*{CRediT authorship contribution statement}

\textbf{M. Schreter-Fleischhacker}:  Conceptualization, Methodology, Software, Validation, Formal analysis, Investigation, Data Curation, Writing - Original Draft, Writing - Review \& Editing, Visualization, Project administration, Funding acquisition
\textbf{N. Much}:  Methodology, Data curation, Software, Writing - Review \& Editing
\textbf{P. Munch}:  Methodology, Data curation, Software, Writing - Review \& Editing
\textbf{M. Kronbichler}: Software, Writing - Review \& Editing
\textbf{W. A. Wall}:  Funding acquisition, Project administration, Resources, Supervision, Writing - Review \& Editing,
\textbf{C. Meier}: Conceptualization, Methodology, Funding acquisition, Project administration, Resources, Supervision, Writing - Original Draft, Writing - Review \& Editing

\section*{Declaration of competing interest}
The authors declare that they have no known competing financial interests or personal relationships that could have appeared
to influence the work reported in this paper.

\section*{Funding sources}
Financial support for this research was provided by the Austrian Science Fund (FWF) through the Schrödinger Fellowship under award number J4577, the European Research Council through the
ERC Starting Grant ExcelAM under award number 101117579 and the Deutsche Forschungsgemeinschaft (DFG, German Research Foundation) under award number 437616465. This support is gratefully acknowledged.

\section*{Acknowledgements}
The authors acknowledge collaboration with Bruno Blais,
Hélène Papillon-Laroche as well as the \texttt{deal.II} community.

\section*{Data availability}

\replaced{The code we use is openly available at \href{\githubRepository}{\githubRepository}.}{Data will be made available on request.}

\appendix

\section{\added{Refined formulations of a consistent level-set transport velocity for a diffuse evaporation-induced velocity jump}}
\label{app:level_set_transport_velocity}

{\color{black}
A critical modeling component of phase change across the liquid-gaseous interface $\Gamma$ lies
in an accurate expression for the level-set transport velocity $\uGamma$ in \myeqref{eq:transport}, which ultimately governs the interface evolution and thus affects the overall dynamics. The interested reader finds an extensive discussion related to the formulation of the level-set transport velocity in our previous contribution~\cite{schreter2024consistent}. Ideally, $\uGamma$ is constructed so that it (i)
predicts the evaporated liquid mass accurately, (ii) forms a continuous field and (iii)
is divergence-free, at least in the near-interface region. However, the determination of 
$\uGamma$ is non-trivial due to the smeared discontinuity in the fluid velocity across the interface in the diffuse one-fluid formulation used in this study. 

We begin with a brief review of our formulations presented in~\cite{schreter2024consistent} for computing the level-set transport velocity under diffuse evaporation-induced velocity jump conditions. These formulations incorporate evaporation-dependent modification of the local fluid velocity and/or projection of the fluid velocity from the liquid or gas phase to the diffuse interface region, 
presented in three distinct variants: \emph{Variant 1} according to Eq.~\eqref{eq:transport_vel_local_continuous}, \emph{variant 2} according to Eq.~\eqref{eq:transport_vel_extension_liquid} and
	\emph{variant 3}, which considers an extension of the fluid velocity from the gaseous end of the interface region $\gasP{\Bx}$ (defined as the projection of a point $\Bx$ along the interface normal $\nGamma$ to the level-set isocontour where $H_\phi(\phi)=0$) 
	\begin{equation}
		\uGamma^{(\text{V3})}(\Bx,t) = \Bu(\gasP{\Bx}(\Bx,t) ) + \frac{\mDot}{\rhoG}\,\nGamma(\Bx,t)
		\qquad
		\text{ for }\Bx \text{ in }\Omega\,.
		\label{eq:transport_vel_extension_gas}
	\end{equation}

Inspired by the hybrid model of Lee et al.~\cite{lee2017direct}, who considered a curvature correction of the evaporative mass flux to compute the level-set transport velocity, we propose two alternative refined formulations of \emph{variant~2} and \emph{variant~3}, enhancing the accuracy:

\emph{Variant 2e} considers an extension of the velocity from the liquid end of the interface region, i.e., from $\liqP{\Bx}$ and additionally performs a curvature correction of this value
\begin{equation}
	\uGamma^{(\text{V2e})}(\Bx,t) = \Bu(\liqP{\Bx}(\Bx,t))\,\frac{\kappa_\Gamma (\Bx,t)}{\liqP{\kappa}(\Bx,t)} + \frac{\mDot}{\rhoL}\,\nGamma(\Bx,t)
	\qquad
	\text{ for }\Bx \text{ in }\Omega\,
	\label{eq:transport_vel_extension_liquid_curv}
\end{equation}
considering the curvature at the interface midplane $\kappa_\Gamma(\Bx)=\kappa(\Bx_\Gamma(\Bx))$ and the curvature at the liquid end of the interface region $\liqP{\kappa}(\Bx)=\kappa(\liqP{\Bx}(\Bx))$. To determine the two curvature values, we again perform an extension algorithm of the interface mean curvature field $\kappa(\Bx)$ from the corresponding level-set isosurfaces. For $\kappa_\Gamma(\Bx)$, we determine the projected point $\Bx_\Gamma(\Bx)$ along the interface normal $\nGamma$ to the level-set isocontour $\phi=0$. Similarly, for $\liqP{\kappa}(\Bx)$, we determine the projected point along the interface normal $\nGamma$ to the level-set isocontour where $H_\phi=1$. 

The concept of the curvature correction factor of \emph{variant~2e} can be similarly applied to \emph{variant~3}. \emph{Variant~3e} considers an extension of the fluid velocity from the gaseous end of the interface region, i.e., from $\gasP{\Bx}$  and additionally performs a curvature correction of this value
\begin{equation}
	\uGamma^{(\text{V3e})}(\Bx, t) = \Bu(\gasP{\Bx}(\Bx, t) )\,\frac{\kappa_\Gamma (\Bx,t)}{\gasP{\kappa}(\Bx,t)} + \frac{\mDot}{\rhoG}\,\nGamma(\Bx,t)
	\qquad
	\text{ for }\Bx \text{ in }\Omega\,
	\label{eq:transport_vel_extension_gas_curv}
\end{equation}
considering the curvature at the interface midplane $\kappa_\Gamma(\Bx)=\kappa(\Bx_\Gamma(\Bx))$  and the curvature at the gaseous end of the interface region $\gasP{\kappa}(\Bx)=\kappa(\gasP{\Bx}(\Bx))$.

To evaluate the refined formulations of the level-set transport velocity, i.e., \emph{variant 2e} and \emph{variant 3e}, we revisit the benchmark example of an evaporating circular shell under isothermal conditions. This example has been introduced in~\cite{schreter2024consistent} and is depicted in Figure~\ref{fig:benchmark_transport_velocity}. At the interface, liquid material evaporates with a spatially and temporally constant evaporative mass flux $\mDot$. Simultaneously, the evaporated volume is compensated by a prescribed inflow velocity on the liquid side of the interface $\bar{u}=\added{(}\mDot/\rhoL\added{)}\added{\cdot(}R_\Gamma/R_\*i\added{)}$ to yield a spatially fixed interface location. 
Thus, we expect the reference solution of ideally a zero level-set transport velocity.
Analytical solutions for the velocity are available for both -- diffuse and sharp models -- and are shown in the top right panel of Figure~\ref{fig:benchmark_transport_velocity}. 

The bottom panel of Figure~\ref{fig:benchmark_transport_velocity} demonstrates the evaluation of the new variants, i.e., variant 2e and variant 3e, compared to variants 2 and 3. In~\cite{schreter2024consistent}, we concluded that the accuracy of predicting the transport velocity for variant 2 is already sufficient for finite interface thicknesses (error $\approx$ $\num{2e-7}$ to $\num{6e-7}$ for the investigated interface thickness values) compared to variant 3 (error $\approx$ $\num{6e-4}$ to $\num{2e-4}$).
The discrepancies in the error behavior arises because, under high evaporation-induced velocity differences between the liquid and gas phases -- due to a significant density ratio and/or large evaporative mass flux -- the transport velocity closely approximates the velocity at the liquid end of the interface region.
However, this can be remedied using the refined formulations for the level-set transport velocity variants 2e and 3e, where perfect agreement with the reference solution can be achieved  for this benchmark case. This is regardless of whether the extension velocity of the liquid phase (variant 2e) or the gas phase (variant 3e) is considered. While this theoretical analysis justifies the improved accuracy achievable with the refined level-set transport formulations, their implementation in a numerical framework are challenging due to the need for additional extension steps, which are computationally expensive, and due to numerical errors in the projected quantities. Given the already high accuracy of variant~2 \eqref{eq:transport_vel_extension_liquid}, we choose not to explore the refined formulations \emph{variant 2e} and \emph{variant 3e}  further in the numerical examples section but leave them here to conclude the theoretical discussion on level-set transport velocities for diffuse phase change models.

\begin{figure}[tb!]
	\centering
	\includegraphics{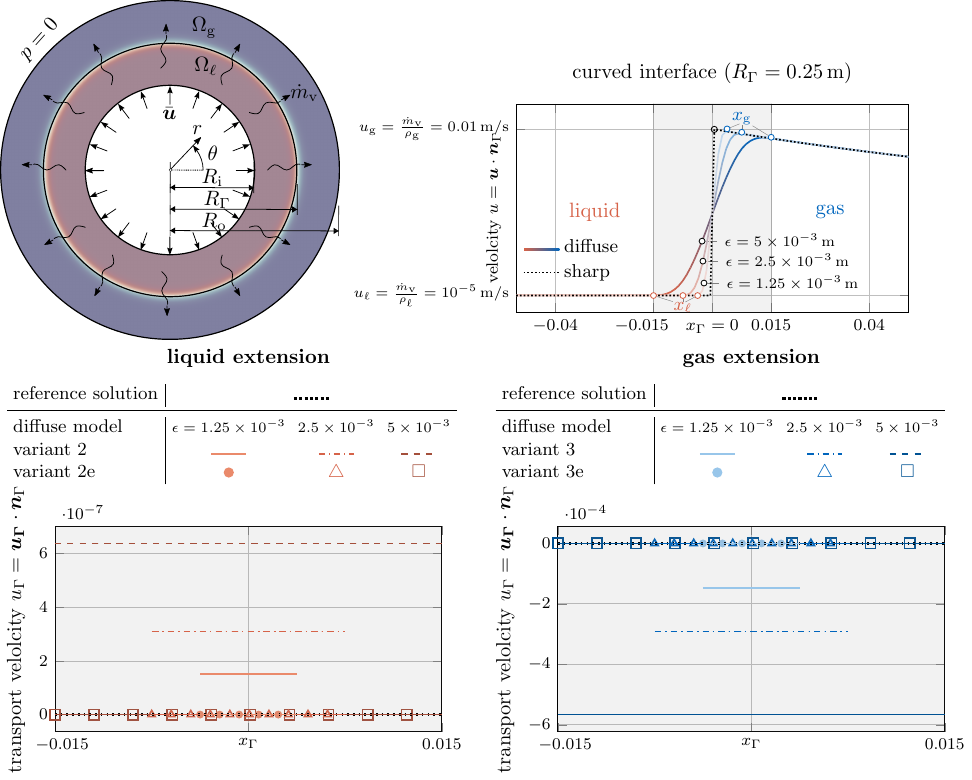}
	\caption{Evaluation of diffuse models based on the analytical solution for the level-set transport velocity of an axisymmetric curved interface subject to evaporation (isothermal conditions): (top left) model setup, taken from~\cite{schreter2024consistent}; (top right) analytical solution for the radial velocity component considering sharp and diffuse modeling of the evaporative velocity jump condition, taken from~\cite{schreter2024consistent}; (bottom) analytical solutions for the level-set transport velocity considering an extension velocity from the liquid end of the interface zone (bottom left) and from the gaseous end of the interface zone (bottom right); Results for \emph{variant 2} and \emph{variant 3} are taken from~\cite{schreter2024consistent}. 
		The parameters are chosen as (SI units): $R_\*i=0.125$, $R_\Gamma=0.25$, $R_\*o=0.375$, $\rhoL=1000$, $\rhoG=1$, $\mDot=0.01$. We note the different scales between the liquid extension and gas extension cases, as well as the overlapping curves for different interface thicknesses in \emph{variant 2e} and \emph{variant 3e}. 
		Both newly introduced variants, i.e., \emph{variant 2e} and \emph{variant 3e}, yield excellent agreement with the  exact level-set transport velocity for arbitrary finite values of the interface thickness. }
	\label{fig:benchmark_transport_velocity}
\end{figure}

}

\section{Numerical framework}
\label{app:numerical_framework}

Detailed information about the numerical framework used to solve the models presented in Sections~\ref{sec:two_phase_flow} and \ref{sec:melt_pool} \replaced{is}{can be} partially found in our previous work~\cite{schreter2024consistent}, which focused exclusively on isothermal conditions. The primary changes in the current work are the inclusion of the energy equation and additional right-hand side contributions. A brief summary of the main aspects of the numerical framework is provided below.

The governing partial differential equations, i.e., \myeqrefs{eq:continuity}{eq:heat_transfer}, and the additional equations for the level-set framework (Section~\ref{sec:level_set})  are spatially discretized using continuous finite elements with Lagrange polynomials as test and trial functions.
The polynomial degrees $k$ are $k_u=2$ for velocity, $k_p=1$ for pressure (to ensure inf-sup stability), $k_\phi=1$ for the level-set, and $k_T=1$ for temperature.  If not stated otherwise, The level-set mesh is refined $\nsub=2$ times, making it twice as fine as the Navier--Stokes mesh, following~\cite{Kronbichler18multiphase}. Numerical integration uses Gaussian quadrature with $(k_i+1)^{dim}$ points, where $i \in {p, u, \phi, T}$.

Time integration employs (semi-)implicit schemes, with a weakly partitioned operator-splitting approach (Algorithm~\ref{algo:meltpooldg}). While each field is propagated fully implicitly, coupling terms (e.g., evaporative dilation rate, recoil pressure, surface tension force, and evaporative cooling) are treated explicitly, introducing a time-step limit. In addition to the capillary time-step limit~\cite{brackbill1992}, the overall time-step limit, influenced by the explicit treatment of additional right-hand side contributions (evaporative dilation rate, recoil pressure, and evaporative cooling), is estimated empirically in subsequent studies due to the lack of studies in this direction.

To achieve high spatial resolution near interfaces, adaptive meshing is employed. 
Linear systems are solved with preconditioned Krylov subspace methods: conjugate-gradient (CG) solvers for symmetric systems and generalized minimal residual method (GMRES) solvers for non-symmetric systems~\cite{saad2003iterative}.
Efficient matrix-free algorithms~\cite{kronbichler2012generic, Kronbichler2019} ensure fast matrix-vector products in iterative solvers, supporting exascale finite-element performance~\cite{kolev2021efficient}. Adaptive mesh refinement and matrix-free methods are utilized from \texttt{deal.II}~\cite{africa2024deal} with parallelized MPI-based domain decomposition. 
In addition, we use and extend the open-source incompressible Navier--Stokes solver \texttt{adaflo}~\cite{Kronbichler18multiphase}, by providing variable material properties and additional right-hand side terms as outlined in Algorithm~\ref{algo:meltpooldg}.

\begin{algorithm}[H]
	\newcommand{\mP}[1]{\textcolor{TUMBlau}{#1}}
	\caption{Overall solution algorithm of the anisothermal two-phase flow model with evaporative phase change (cf. Section~\ref{sec:two_phase_flow}) and the multi-phase melt pool model (extensions marked in \mP{blue}, cf. Section~\ref{sec:melt_pool}).}\label{algo:meltpooldg}
	$\Bu\gets\Bu^{(0)},p\gets p^{(0)},T\gets T^{(0)},\phi\gets \phi^{(0)}$ \Comment*[r]{set initial conditions at $t=t^{(0)}$}
	\For{$t\gets t^{\mathrm{(start)}}$ \KwTo $t^{\mathrm{(end)}}$}{
		\tcc{\underline{step 1:~compute predictors of the primary variables}}
		$\Bu, p, T, \phi\gets \texttt{extrapolate}(\Bu^{(n)}, \; p^{(n)}, \; T^{(n)}, \; \phi^{(n)},\;\Bu^{(n-1)}, \; p^{(n-1)}, \; T^{(n-1)}, \; \phi^{(n-1)}, \; ...)$\;
		\tcc{\underline{step 2:~compute level-set and geometric quantities}}
		$\phi,\;\nGamma,\;\kappa \gets \texttt{level\_set\_solver}(\phi,\;\uGamma, \mDot)$  (cf. algorithm in~\cite{schreter2024consistent})\;
		\tcc{\underline{step 3:~solve energy equation}}			
		$s \gets \evaporCooling\left(\phi,T\right)~\eqref{eq:evaporDil}\mP{ +\laserFlux\left(\nGamma,\phi\right)}$~\mP{\eqref{eq:laser_gauss}} \Comment*[r]{laser heat source, evaporative heat loss} 
		$T\gets \texttt{heat\_solver}(T, \phi, \Bu, s)$\\
		\tcc{\underline{step 4:~solve incompressible Navier--Stokes equations}}
		$\eff{\rho}\gets\rho(\phi\mP{,T}), \eff{\mu}\gets\mu(\phi\mP{,T}), \mP{K\gets K(\phi,T)}$\Comment*[r]{compute effective material properties}
		$\evaporDilationRate \gets \evaporDilationRate (\mDot, \phi)$~\eqref{eq:evaporDil},\
		$\boldsymbol{f} \gets \surfaceTensionForce(\kappa,\nGamma,\phi,T)~\eqref{eq:surface_tension} \mP{ + \recoilPressureForce(\nGamma,\phi,T)~\eqref{eq:hybrid_recoil_pressure} + \DarcyDampingForce(\phi,T)~\eqref{eq:darcy_damping_force}}$ \\\hfill\Comment{compute force and flux contributions}
		$\Bu,p \gets \texttt{navier\_stokes\_solver}(\eff{\rho},\eff{\mu}, K, \evaporDilationRate, \boldsymbol{f})$ (cf. algorithm in~\cite{Kronbichler18multiphase})
	}
\end{algorithm}

\section{Reinterpretation of the Tanasawa Model through a penalty method approach}
	\label{remark_1}
	
	\newcommand{\testPhi}{\delta T}
	\newcommand{\testPhih}{\delta T^h}
	\newcommand{\penaltyParam}{\eta_\*{p}}
	
	\added{We demonstrate that} using the Tanasawa model for the evaporative mass flux according to \myeqref{eq:mDotTanasawa} to model evaporative cooling across a liquid--vapor interface $\Gamma$ in the energy equation \eqref{eq:heat_transfer} \replaced{constitutes}{can be interpreted as} a penalty method approach that enforces the boiling temperature $\Tv$ at the liquid--vapor interface.
	Let us consider the strong form of the energy equation 
	\begin{equation}
		\rhoCpEff\left(\fracPartial{T}{t} + \boldsymbol{u}\cdot\nabla T\right) = \nabla\cdot{\left(\kEff\,\nabla T\right)}
		\quad \text{ in }\Omega\times[0,\tEnd]\,
		\label{eq:heat_penalty}
	\end{equation}
	subject to the constraint
	\begin{equation}
		T=\Tv
		\quad \text{ on }\Gamma\times[0,\tEnd]\,.
		\label{eq:heat_dirichlet_constraint}
	\end{equation}
	For demonstration purposes, in~\myeqref{eq:heat_penalty}  we neglect the additional fluxes $s$ compared to those included in \myeqref{eq:heat_transfer}. 
	The weak form of \myeqref{eq:heat_penalty} is obtained by multiplying with the weighting functions for the temperature field, denoted as $\testPhi$,
	and	integration over the spatial domain: 
	\begin{equation}
		\label{eq:weakFormHeat}
		\underbrace{\BiLi{\testPhi}{\rhoCpEff\left(\fracPartial{T}{t} + \boldsymbol{u}\cdot\nabla T\right) - \nabla\cdot{\left(\kEff\,\nabla T\right)}}}_{\delta W_\Omega} = 0 \quad \,. 
	\end{equation}
	We formulate the penalty potential to enforce the Dirichlet constraint of \myeqref{eq:heat_dirichlet_constraint} as
	\begin{equation}
		\Pi_\*p = \int_\Gamma \frac{\penaltyParam}{2}\left(T-\Tv\right)^2
	\end{equation}
	with the penalty parameter $\penaltyParam$. Incorporating the penalty term into the weak form \eqref{eq:weakFormHeat} results in
	\begin{equation}
		0 = \delta W_\Omega + \BiLiPure{\testPhi}{\penaltyParam\left(T-\Tv\right)}{\Gamma}\,.
		\label{eq:weakPlusPenalty}
	\end{equation}
	We consider a regularized interface flux formulation for the surface integral of the penalty term in \myeqref{eq:weakPlusPenalty} using the parameter-scaled delta function $\deltaRhoCp$ and obtain the regularized expression
	\begin{equation}
		0 = \delta W_\Omega + \BiLiPure{\testPhi}{\penaltyParam\left(T-\Tv\right)\deltaRhoCp}{\Omega}\,.
		\label{eq:penalty_virtual_work}
	\end{equation}
	\replaced{Comparison of}{By comparing} \myeqref{eq:penalty_virtual_work} to the evaporative cooling term in \myeqref{eq:vapor_heat_loss} \replaced{reveals }{it becomes apparent} that the penalty parameter \replaced{has a physical interpretation}{can be physically interpreted} as $\penaltyParam=\evaporationHeatTransferCoefficient\,\hv$.
	
	Since we solve the energy equation \eqref{eq:heat_penalty} (or \myeqref{eq:heat_transfer}), the resulting evaporative cooling flux \eqref{eq:penalty_virtual_work} is independent of the penalty parameter, provided the latter is not chosen unrealistically low. The penalty parameter $\penaltyParam$ primarily influences the extent to which the temperature deviates from the target interface temperature $\Tv$. However, the magnitude of this temperature violation scales directly with the penalty parameter.  
	If the penalty parameter is sufficiently high, the model imposes a strong penalty for temperature deviations from $\Tv$, effectively enforcing the Dirichlet condition. Nevertheless, if $\penaltyParam$ is too high, the resulting system of equations \deleted{may} become\added{s} ill-conditioned, leading to potential numerical difficulties.
	
\section{Derivation of the recoil pressure model by Anisimov from a macroscopic continuum perspective}
\label{app:recoil_pressure}
\begin{figure}[tbp!]
	\includegraphics{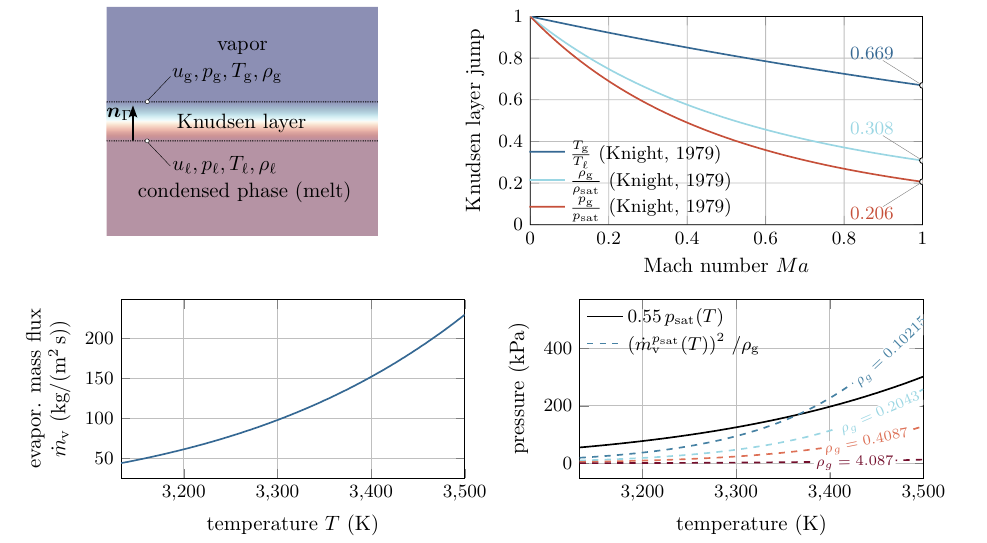}
	\caption{(top left) sketch of the Knudsen layer; (top right) Illustration of the jump conditions for the temperature $\gasP{T}/\liqP{T}$, the density $\rhoG/\rhoL$ and the pressure $\gasP{p}/\liqP{p}$ over the Knudsen layer according to~\cite{knight1979theoretical}, Eq. 6 for a monoatomic gas with $\gamma_v=5/3$; (bottom left) evaporative mass flux \eqref{eq:net_evaporation_rate_mach1} for $\Mach=1$ using the parameters of Table~\ref{tab:param_ti64}; (bottom right) illustration of the evaporation-induced pressure contribution for $\Mach=1$ 
	using the parameters of Table~\ref{tab:param_ti64}.}
	\label{fig:knudsen}
\end{figure}

Inside the Knudsen layer (top left panel of Figure~\ref{fig:knudsen}), which is formed directly between the condensed matter and the vapor phase with a typical thickness of a few molecular mean free paths, usually non-equilibrium effects are observed following the kinetic gas theory based on Boltzmann's equation~\cite{knight1979theoretical,klassen2014evaporation}. 
From a macroscopic viewpoint, the Knudsen layer represents a contact discontinuity, where rapid vapor expansion and gas
compression dynamics lead to discontinuities in the flow and temperature fields. 
The resulting jumps in the temperature, the density and the pressure over the Knudsen layer \replaced{are described}{can be modeled} by analytical 
macroscopic jump conditions depending on the Mach number $\Mach$ of the vapor phase\added{, as} derived by Knight~\cite{knight1979theoretical}\deleted{,} and illustrated in the top right panel of Figure~\ref{fig:knudsen}. The macroscopic jump conditions for temperature, density and pressure read
\def\TL{\liqP{T}}
\def\TG{\gasP{T}}
\def\mG{\gasP{m}}
\def\erfc{\text{erfc}}
\begin{align}
	\frac{\gasP{T}}{\liqP{T}} &= \left(\sqrt{1+\left(\pi\,\frac{(\gamma_v-1)\,\gasP{m}}{2\,\left(\gamma_v+1\right)}\right)^2}-\sqrt{\pi}\frac{(\gamma_v-1)\,\gasP{m}}{2\,\left(\gamma_v+1\right)}\right)^2\,,
	\label{eq:jumpTemperature}\\
	\frac{\gasP{\rho}}{\rhoSat(\TL)} &= \sqrt{\frac{\TL}{\TG}}\left(\frac{2\,\mG^2+1}{2}\exp\left(\mG^2\right)\erfc(\mG)-\frac{\mG}{\sqrt{\pi}}\right) + \frac{\TL}{2\,\TG} \left(1-\sqrt{\pi}\mG\exp(\mG^2)\erfc(\mG)\right)\,,	\label{eq:jumpDensity}\\
	\frac{\gasP{p}}{\pSat(\TL)} &= \frac{\gasP{\rho}\,\TG}{\rhoSat(\TL)\,\TL}\,
	\label{eq:jumpPressure}
\end{align}
with the specific heat ratio for monoatomic gases $\gamma_v=5/3$~\cite{knight1979theoretical}  and the dimensionless velocity $\mG=\sqrt{\gamma_v/2}\Mach$.
%
Based on the macroscopic jump conditions, we derive the formulation for the evaporation-induced recoil pressure by Anisimov et al.~\cite{anisimov1995instabilities} in the following.
We depart from the Hertz-Knudsen formula for the evaporative mass flux
\begin{equation}
	\mDot^+(\TL) = \pSat(\liqP{T})\sqrt{\frac{M}{2\,\pi\,R\,\liqP{T}}}\,.
	\label{eq:mDotHertzKnudsenFull}
\end{equation}
with the universal gas constant $R$, the molar mass $M$ and the Clausius-Clapeyron relation for the saturated vapor pressure at temperature of the condensed surface $\liqP{T}$:
\begin{equation}
		\pSat(\liqP{T}) = \pAmbient \exp\left(\frac{\hv M}{R \Tv} \left(1-\frac{\Tv}{\liqP{T}}\right)\right)\,.
		\label{eq:ClausiusClapeyron}
\end{equation}
The condensation mass flux reads
\begin{equation}
	\mDot^-(\TG) = \gasP{p}\sqrt{\frac{M}{2\pi \gasP{T} R}}\,.
	\label{eq:condensationHertzKnudsenFull}
\end{equation}
From the evaporative mass flux and the condensation flux, the net evaporation flux is obtained as
\begin{align}
 \mDot(\TL) &= \mDot^+ -\mDot^- \\
            &= \varphi\,\mDot^+ 
 \\ &=\varphi\,\pSat{}(\TL)\sqrt{\frac{M}{2\,\pi\,R\,\TL}}\,.
	\label{eq:mDotHertzKnudsen}
\end{align} 
The coefficient
\begin{equation}
\varphi = \sqrt{2\pi\gamma_v} \Mach \frac{\rhoG}{\rhoSat(\liqP{T})} \sqrt{\frac{\gasP{T}}{\liqP{T}}} 
\end{equation}
takes into account the reduction of the evaporation flux due to condensation. 
Since the analytical expressions for the jumps in density \eqref{eq:jumpDensity} and temperature \eqref{eq:jumpTemperature} depend solely on the Mach number, the coefficient $\varphi$ is also only dependent on the Mach number. The maximum value of $\varphi=0.82$ is obtained at $\Mach=1$, which 
results in the net evaporation rate 
\begin{equation}
	\mDotPSat(\Tl)  = 0.82\,\pSat(\liqP{T})\sqrt{\frac{1}{2\,\pi\,R_s\,\liqP{T}}}\,.
	\label{eq:net_evaporation_rate_mach1}
\end{equation}
It is illustrated in the bottom left panel of Figure~\ref{fig:knudsen} for the parameters in Table~\ref{tab:param_ti64} representing \tiSixFour{}.
This model for the evaporative mass flux is typically employed in computational melt pool models and also adopted in this work. \added{We note that the assumption of a constant Mach number of~1 corresponds to near-vacuum evaporation~\cite{klassen2014evaporation} and represents a simplification in the presented evaporation model. While this approach captures key evaporation dynamics, it cannot account for pressure-dependent scenarios where vapor confinement inside deep keyholes increases local pressure, potentially reducing the net evaporation rate.\\} If we calculate the evaporation-induced velocity and pressure jumps across the liquid--vapor interface 
from the conservation of mass and momentum (Rankine-Hugoniot conditions), we obtain
 \begin{align}
	\label{eq:velocity_jump}
	\left(\liqP{\boldsymbol{u}} - \gasP{\boldsymbol{u}}\right)\cdot \nGamma  &= \Big(\frac{1}{\rhoL}-\frac{1}{\rhoG}\Big)\,\mDotPSat
	\underbrace{\approx}_{\rhoG \ll \rhoL} -\frac{\mDotPSat}{\rhoG}\,, \\
	\label{eq:pressure_jump}
	\liqP{p} - \gasP{p} &= \Big(\frac{1}{\rhoG}-\frac{1}{\rhoL}\Big)\,\left(\mDotPSat\right)^2 \underbrace{\approx}_{\rhoG \ll \rhoL} \frac{\left(\mDotPSat\right)^2}{\rhoG}\,.
\end{align}
Considering the macroscopic jump conditions according to Figure~\ref{fig:knudsen}, we obtain the pressure and density at the vapor surface for $Ma=1$ as $\gasP{p}(T_g)=0.206\,\pSat(\TL)$ and $\rhoG(\TL)=0.308\rhoL(T_l)$, respectively, whereas $\rhoL=\pSat(\TL)/(\TL\,R/M)$ follows from the ideal gas law. After insertion of these expressions as well as \myeqref{eq:net_evaporation_rate_mach1} and \myeqref{eq:ClausiusClapeyron}  into \myeqref{eq:pressure_jump} and rearrangement, an expression for the pressure at the condensed surface is obtained as 
\begin{align}
	\label{eq:pressure_liquid_side}
	\underbrace{	p_{\text{recoil}}(\TL)}_{\text{recoil pressure}} &= \gasP{p}(\liqP{T}) +  \frac{\mDotPSat(\TL)^2}{\rhoG}  \nonumber \\ 
	&= \underbrace{0.206\,\pSat(\liqP{T})}_{\gasP{p}} + \frac{1}{0.308\rhoSat(\TL)}\,\left(0.82\,\pSat(\liqP{T})\sqrt{\frac{M}{2\,\pi\,R\,\liqP{T}}}\right)^2 \nonumber \\
	&=0.206\,\pSat(\TL) + \underbrace{\frac{\cancel{R\,\liqP{T}}}{0.308\,\cancel{M}\cancel{\pSat(\TL)}}\,0.82^2\pSat\cancel{^2}(\TL) \frac{\cancel{M}}{2\,\pi\,\cancel{R\,\liqP{T}}}}_{0.347\pSat} \nonumber\\ 
	&=0.55\,\pSat(\TL) \nonumber\\
	&= 0.55\,\pAmbient\,\exp\left(-\frac{\hv}{R}\left(\frac{1}{\TL}-\frac{1}{\Tv}\right)\right)\,.
\end{align}
representing the expression for the recoil pressure model by Anisimov et al.~\cite{anisimov1995instabilities}.

In the bottom right panel of Figure~\ref{fig:knudsen}, \myeqrefs{eq:pressure_jump}{eq:pressure_liquid_side} are illustrated, where for  \eqref{eq:pressure_jump} we consider different temperature- and pressure-independent vapor phase densities. It \replaced{follows}{can be seen} that considering an incompressible flow framework with evaporative phase change, the inherently induced pressure jump condition at the liquid--vapor interface (\myeqref{eq:pressure_jump} or similar \myeqref{eq:recoilPressureEvaporDil}) together with a temperature- and pressure-independent vapor density cannot be calibrated such that the recoil pressure model by Anisimov is reobtained. This was the motivation for introducing the hybrid recoil pressure force in Section~\ref{sec:hybrid_evapor}.

\section{\added{Computational effort for the simulations of stationary laser illumination}}
\label{app:comp_effort}

\added{Table~\ref{tab:comp_effort} summarizes the computational effort for simulating stationary laser illumination in 2D (cf. Section~\ref{sec:melt_pool_2d}) and in 3D (cf. Section~\ref{sec:melt_pool_3d}), representing the most demanding simulations in this numerical study. Notably, using level-set transport velocity variant 2 consumes a significant portion of the total runtime, highlighting the need for performance optimization in large-scale simulations.}

\begin{table}[tbp!]
	\caption{\added{Distribution of the computational effort of 2D and 3D stationary laser illumination simulations.}}
	\label{tab:comp_effort}
	\centering
	\begin{tabular}{p{6cm}||c|c||c} 
		\toprule
		 & \multicolumn{3}{c}{stationary laser illumination} \\
		 & \multicolumn{2}{c}{2D} & 3D\\
		$\uGamma$  & variant 1 & variant 2 & variant 1 \\
		\midrule
		CPU & AMD Ryzen Thread- & \multicolumn{2}{c}{Intel Xeon Gold 6230}\\
		&ripper PRO 3995WX\\
		number of cores  & 62 & 120 & 240\\
		overall run time   & \SI{11}{hours} & \SI{76}{hours} & \SI{43}{hours}\\
		avg. number of DoFs ($\phi + T + p + \Bu$)  & \num{981513}  & \num{1182930} &  \num{4619605} \\
		avg. number of  finite elements  & \num{68492}& \num{105873} & \num{757013}\\
		total number of time steps   &  \num{e5} &  \num{e5} & \num{e5} \\
		\midrule
		\midrule
		task & \multicolumn{3}{c}{relative effort} \\
		\midrule
		- Navier--Stokes  &  \SI{24}{\%} &  \SI{35}{\%} & \SI{57}{\%}  \\
		- heat transfer  & \SI{13}{\%}  &  \SI{1}{\%} & \SI{13}{\%}  \\
		- level set (advection, reinitialization, normal vector, curvature)  & \SI{58}{\%} &  \SI{4}{\%} &  25\,\% \\
		- level-set transport velocity (note: not optimized!)  &  \SI{2}{\%}  &  \SI{60}{\%} & 3\,\%\\
		- other (adaptive mesh refinement, output, initial conditions, etc.)  & $< 3\,\%$ & $< 1\,\%$ & $< 2\,\%$ \\
		\bottomrule
	\end{tabular}
\end{table}

\bibliographystyle{elsarticle-num} 
\newpage
\bibliography{meltpool.bib}

\end{document}